Transmission Electron Microscopy Study of the Morphology of Ices Composed of $H_2O$, $CO_2$, and CO on Refractory Grains

Short title: TEM study of Morphology of Ices on Refractory Grains

Akira Kouchi[1], Masashi Tsuge[1], Tetsuya Hama[2], Yasuhiro Oba[1], Satoshi Okuzumi[3], Sin-iti Sirono[4], Munetake Momose[5], Naoki Nakatani[6], Kenji Furuya[7], Takashi Shimonishi[8], Tomoya Yamazaki[1], Hiroshi Hidaka[1], Yuki Kimura[1], Ken-ichiro Murata[1], Kazuyuki Fujita[1], Shunichi Nakatsubo[9], Shogo Tachibana[10], and Naoki Watanabe[1]

[1] Institute of Low-Temperature Science, Hokkaido University, Sapporo, Hokkaido 060-0819, Japan; tsuge@lowtem.hokudai.ac.jp

[2] Komaba Institute for Science, The University of Tokyo, Meguro, Tokyo 153-8902, Japan

[3] Department of Earth and Planetary Sciences, Tokyo Institute of Technology, Meguro, Tokyo 152-8551, Japan

[4] Earth and Environmental Sciences, Graduate School of Environmental Studies, Nagoya University, Nagoya, Aichi 464-8601, Japan

[5] College of Science, Ibaraki University, Mito, Ibaraki 310-8512, Japan

[6] Department of Chemistry, Graduate School of Science and Engineering, Tokyo Metropolitan University, Hachioji, Tokyo 192-0397, Japan

[7] National Astronomical Observatory of Japan, Osawa 2-21-1, Mitaka, Tokyo 181-8588, Japan

[8] Center for Transdisciplinary Research, Niigata University, Niigata, Niigata 950-2181, Japan

[9] Institute of Space and Astronautical Science, Japan Aerospace Exploration Agency, Sagamihara, Kanagawa 252-5210, Japan

[10] UTokyo Organization for Planetary and Space Science/Department of Earth and Planetary Science, University of Tokyo, Hongo, Tokyo 113-0033, Japan

Abstract

It has been implicitly assumed that ices on grains in molecular clouds and proto-planetary disks are formed by homogeneous layers regardless of their composition or crystallinity. To verify this assumption, we observed the $H_2O$ deposition onto refractory substrates and the crystallization of amorphous ices ($H_2O$, $CO_2$, and $CO$) using an ultra-high-vacuum transmission electron microscope. In the $H_2O$-deposition experiments, we found that three-dimensional islands of crystalline ice ($I_c$) were formed at temperatures above 130 K. The crystallization experiments showed that uniform thin films of amorphous CO and $H_2O$ became three-dimensional islands of polyhedral crystals; amorphous $CO_2$, on the other hand, became a thin film of nano-crystalline $CO_2$ covering the amorphous $H_2O$. Our observations show that crystal morphologies strongly depend not only on the ice composition, but also on the substrate. Using experimental data concerning the crystallinity of deposited ices and the crystallization timescale of amorphous ices, we illustrated the criteria for ice crystallinity in space and outlined the macroscopic morphology of icy grains in molecular clouds as follows: amorphous $H_2O$ covered the refractory grain uniformly, $CO_2$ nano-crystals were embedded in the amorphous $H_2O$, and a polyhedral CO crystal was attached to the amorphous $H_2O$. Furthermore, a change in the grain morphology in a proto-planetary disk is shown. These results have important implications for the chemical evolution of molecules, non-thermal desorption, collision of icy grains, and sintering.

Subject Keywords

Experimental techniques (2078); Ice formation (2092); Laboratory astrophysics (2004)

## 1. Introduction

Collision and subsequent sticking or fragmentation of grains or grain aggregates is among the most important of elemental processes for planetesimal formation in proto-planetary disks. Much theoretical research has simply assumed spherical grains for dust aggregation (e.g., Dominik & Tielens 1997; Wada et al. 2007); these studies showed that the kinetic energy of grains is dissipated by their rolling, sliding, and twisting motions, and that sticking of grains or grain aggregates occurs at velocities lower than a critical collision velocity $v_c$. Many collision experiments have been also performed using spherical samples, e.g., silica glass (e.g., Blum & Wurm 2000) or ice crystals (e.g., Gundlach & Blum 2015; Gärtner et al. 2017; and references therein). These experiments have confirmed that sticking of grains or grain aggregates occurred at velocities below $v_c$. However, Poppe et al. (2000) demonstrated that sticking or fragmentation occurred randomly regardless of collision velocities when irregularly shaped grains were used in collision experiments; this suggests that the collision-sticking models developed for spherical grains might not be applicable to the collision of irregularly shaped grains.

Are grains in proto-planetary disks spherical (ellipsoidal) or irregularly shaped? In the Solar System, we can partially answer this question by observing chondritic porous (CP) interplanetary dust particles (IDP) and cometary dust. Typical CP-IDP is an aggregate of μm to sub-μm sized irregularly-shaped anhydrous silicates (e.g., Brownlee 1985; Flynn et al. 2013; Bradley 2014). Close up observation of dust particles from comet 67P/Churyumov-Gerasimenko also showed that porous and fluffy dust are major components of cometary particles (e.g., Hilchenbach et al. 2016; Güttler et al. 2019). Because we could not obtain any information concerning the shapes of ices from the observation of IDP and collected cometary particles, we have therefore assumed that icy grains are ellipsoids regardless of composition and crystallinity. Is this picture correct? We attempt to answer this question by focusing on ices in this study.

In low-temperature molecular clouds, icy mantles are composed of $H_2O$, CO, $CO_2$, $NH_3$, and $CH_3OH$ (e.g., Gibb et al. 2004; Boogert et al. 2015), and $H_2O$ ice is generally considered to be amorphous (e.g., Hagen et al. 1981; Smith et al. 1989). When a protostar and its proto-planetary disk form, crystallization of amorphous ice and subsequent sublimation of respective molecules occur depending on the heliocentric distance from the protostar. The spherical layered structure of icy grains has been assumed in most previous studies discussing their physical and chemical evolution in proto-planetary disks. Ehrenfreund et al. (1998) proposed that the spherical structures of icy grains toward massive protostars consisted of a silicate core, an inner polar-ice mantle, and an outer apolar-ice mantle. Pontoppidan et al. (2008), Boogert et al. (2015), and Öberg (2016) suggested a similar layered structure of ice in molecular clouds; this consisted of $H_2O$-rich ice in the interior, a mixture of $CO_2$ and CO ice in the middle, and CO-rich ice on the exterior. Similar layered-ice structures have also been adopted in some recent models for proto-planetary dust growth (Musiolik et al. 2016; Pinilla et al. 2017; Okuzumi & Tazaki 2019).

In the previous studies named above, spherically-shaped icy grains have been assumed regardless of crystallinity. For amorphous ices, this assumption may be valid because the self-diffusion coefficients of their respective molecules are small at low temperatures, resulting in the formation of thin-layered amorphous ice mantle. For crystalline ices, on the other hand, this assumption may not necessarily be correct; crystallization of amorphous ice requires rearrangement of molecules through diffusion and may result in morphological changes to a thin homogeneous layer.

To clarify the crystallinity of ices, we must know the formation condition of amorphous ices during deposition, as well as their preservation conditions, as demonstrated by Kouchi et al. (1994). The crystallinities of $H_2O$ (Kouchi et al. 1994) and CO (Kouchi et al. 2021) ices during deposition have been investigated using reflection high-energy electron diffraction (RHEED) and transmission electron microscopy (TEM), respectively. Amorphous $H_2O$ (a-$H_2O$) and crystalline CO (α-CO) were found to have been formed under the conditions found in molecular clouds; however, Kouchi et al. (1994) used polycrystalline ice $I_c$ substrate for the $H_2O$-deposition experiments. It is highly desirable to use astrophysically relevant substrates instead of ice $I_c$; these include amorphous silicate, amorphous carbon, and organic materials. For the crystallinity of $CO_2$, although Kouchi et al. (2020) observed the deposition process of $CO_2$ on a-$H_2O$, they did not discuss the crystallinity of $CO_2$ in molecular clouds.

We use the prefix "a-" as an abbreviation for amorphous in the present paper. After "a-" follows the chemical composition (e.g., a-C or a-Si:H). The term "amorphous solid water" is often used for vapor-deposited a-$H_2O$; however, we do not use this term in the present paper to clarify chemical composition of amorphous materials and to avoid misunderstanding. Crystalline polymorphs are usually distinguished using Greek letters (e.g., α-CO, β-quartz) or Roman numerals (e.g., $CO_2$ I, ice VI). In addition, the word "ices" is used to denote a solid at low temperatures (including $H_2O$, $CO_2$, and CO), regardless of crystallinity.

The crystallization of a-$H_2O$ has been extensively studied using X-ray diffraction (Dowell & Rinfret 1960), infrared spectroscopy (e.g., Hagen et al.; 1981; Schmitt et al. 1989), temperature-programmed desorption (TPD) spectroscopy (e.g., Kouchi 1987; Smith et al. 1997; Chakarov & Kasemo 1998; Dohnálek et al. 1999; Fraser et al. 2001), and RHEED (Kouchi 1990). Crystallization of a-CO has been also investigated using RHEED and TPD (Kouchi 1990) and TEM (Kouchi et al.

2021), and that of a-$CO_2$ has been investigated using IR (Escribano et al. 2013; Gerrakines & Hudson 2015; Baratta & Palumbo 2017; He & Vidali 2018; Tsuge et al. 2020). Although Kouchi et al. (1994) theoretically analyzed the crystallization of a-$H_2O$, they did not consider the morphology of a newly formed crystal.

In spite of previous crystallization studies of various amorphous ices, direct observations of newly formed crystals focusing on both crystal structure and morphology have been limited to the following studies on $H_2O$. Jenniskens & Blake (1996) observed crystallization of a-$H_2O$ deposited at 14 K on an a-C film using a TEM and showed that crystalline ices are not uniform films but three-dimensional islands (Fig. 4 of Jenniskens & Blake 1996). By using TEM, Tachibana et al. (2017) observed a similar behavior in the crystallization of a-$H_2O$ on a-Si; i.e., crystalline ice $I_c$ formed at around 145 K from a-$H_2O$ on a-Si also took the form of three-dimensional islands. These results suggest the possibility that crystalline $H_2O$ ice grains are not uniform ellipsoids but polyhedrons. In molecular clouds and proto-planetary disks, a-$H_2O$ could form on various substrates: a-silicate (a-$Mg_2SiO_4$, a-$MgSiO_3$), a-C, and organic materials. However, the only astrophysically relevant substrate used in previous studies was a-C by Jenniskens & Blake (1994, 1996). Collings et al. (2015) speculated from the measurement of infrared spectra of a-$H_2O$ on amorphous silica (a-$SiO_2$) that islands of a-$H_2O$ might be formed at temperatures higher than 40 K when thickness of amorphous $H_2O$ was smaller than one monolayer. Note that the use of a-$SiO_2$ substrate is not adequate as a model of amorphous silicates in molecular clouds, because it does not include MgO and FeO. Since the crystal's morphology on the substrate was determined not only by its equilibrium form but also by the interfacial energy between it and the substrate, it is highly desirable to perform deposition and crystallization experiments using astrophysically relevant substrates. Kouchi et al. (2020, 2021) observed the morphology of α-CO on a-$H_2O$ using TEM and found that the morphology of α-CO was not ellipsoidal with an onion-like structure (as previously assumed), but rather a polyhedral crystal attached to a-$H_2O$.

In the present study, we systematically observed the morphology of ice $I_c$ deposited on astrophysically relevant substrates (organics, a-silicate, and a-C), as well as the morphological changes of a-$H_2O$, a-$CO_2$, and a-CO during crystallization on astrophysically relevant substrates (i.e., organics, a-silicate, and a-C for the deposition of a-$H_2O$; a-$H_2O$ and other ices for a-CO and a-$CO_2$) using ultra-high-vacuum TEM. We also discussed the astrophysical implications of these experimental results.

In the following section, we briefly describe the experimental techniques and protocols. Experimental results on the deposition of $H_2O$ onto refractory substrates and on crystallization of amorphous ices are presented in section 3. In section 4, based on TEM observation, we discuss the formation of ices in molecular clouds, and evolution of icy grains in proto-planetary discs. The astrochemical and astrophysical implications of these results are discussed in section 5, including nonthermal desorption of molecules, collision and sticking of icy grains, and sintering of icy grains.

## 2. Experimental

*2.1. Ultra-high-vacuum transmission electron microscope*

We developed a 200-kV ultra-high-vacuum transmission electron microscope (UHV-TEM) (JEM-2100VL, JEOL) for *in situ* deposition and observation of ices following Kondo et al. (1991), as described in our recent papers (Kouchi et al. 2016; Tachibana et al. 2017; Kouchi et al. 2020). Briefly, a microscopic column is evacuated using five sputter ion pumps and two Ti-sublimation pumps. The pressure near the specimen is measured by a nude ionization gauge placed between the sample chamber and the ion pump. Because the sample was surrounded by a liquid nitrogen shroud, the real pressure near the specimen was lower than the measured pressure.

We used a commercially available liquid He cooling holder (ULTST, Gatan) for specimen cooling. Because the temperature sensor's position is not the same as that of the specimen, the temperature of the specimen differs from that of the sensor. The temperature of the specimen is calibrated using the vapor pressures of crystalline Ne, CO, Ar, and $CO_2$. The temperature difference between the specimen and the temperature sensor at low temperatures, such as 10–20 K, is around 3 K (with the temperature of the specimen being higher than that of the sensor), and at temperatures above 30 K, almost no difference occurs. The errors in temperature measurement at 10–30 K and > 30 K are about ± 1.2 K and ± 0.5 K, respectively. One of the three ICF 70 ports directed at the sample surface with an incident angle of 55° is used for gas deposition, with a variable-leak valve connected to a 0.4-mm inner-diameter Ti-gas inlet tube. A small gas-mixing system is attached to the variable-leak valve.

*2.2. Refractory substrates*

A non-porous a-Si film of thickness 5 nm sputtered onto the Si single-crystal grid (SiMPore Inc. US100-A05Q33) was used as a substrate for the deposition of ices, organic matter, and a-silicate for the following reasons: 1) because the thermal conductivity of the Si single crystal at 10–200 K is larger than $10^3$ W m$^{-1}$ K$^{-1}$ (Glassbrenner and Slack, 1964), the a-Si film is deposited very firmly on the single Si crystal; 2) the TEM contrast of 5-nm-thick a-Si is very weak; and 3) the edge of a Si single crystal could be used as a standard for camera-length calibration in electron diffraction. We observed the non-porous a-Si film using a high-resolution field emission TEM (JEM-2100F, JEOL) and observed neither pores nor cracks (as shown in Fig. 1a).

An organic refractory material (organics) was made using the method of Piani et al. (2017). A gas mixture consisting of $H_2O:CH_3OH:NH_3$ = 6:4:1 was deposited onto the a-Si substrate and simultaneously irradiated with UV-rays at around 10 K using the PICACHU setup. After deposition, the substrate was warmed up to room temperature and organic refractory materials remained. Although the organic refractory material was not of uniform thickness (see Figs. 3e, f of Piani et al. 2017), we observed the uniform part in the present study (Fig. 1b). The electron-diffraction pattern shows that organic refractory material is amorphous.

An a-silicate was deposited onto the a-Si substrate at room temperature via magnetron sputtering of a $Mg_2SiO_4$ polycrystalline target. The thickness measured by the quartz microbalance was 10 nm and the density of a-silicate was assumed to be the same as that of crystalline $Mg_2SiO_4$ (forsterite). TEM images and electron-diffraction patterns show that the film is reasonably uniform and amorphous (Fig. 1c), and the atomic composition measured using EDS equipped with JEM-2100F was Mg:Si:O = 33.2:15.6:51.2 (atm%). Hereafter, we call the a-silicate a-$Mg_2SiO_{3.3}$.

For the a-C substrate, we used commercially available C-flat with a thickness of 20 nm (Protochip Inc.) (Quispe et al. 2007). As shown in Fig. 1d, the film is amorphous and has many small holes, and there are some tiny graphitic grains on the a-C film.

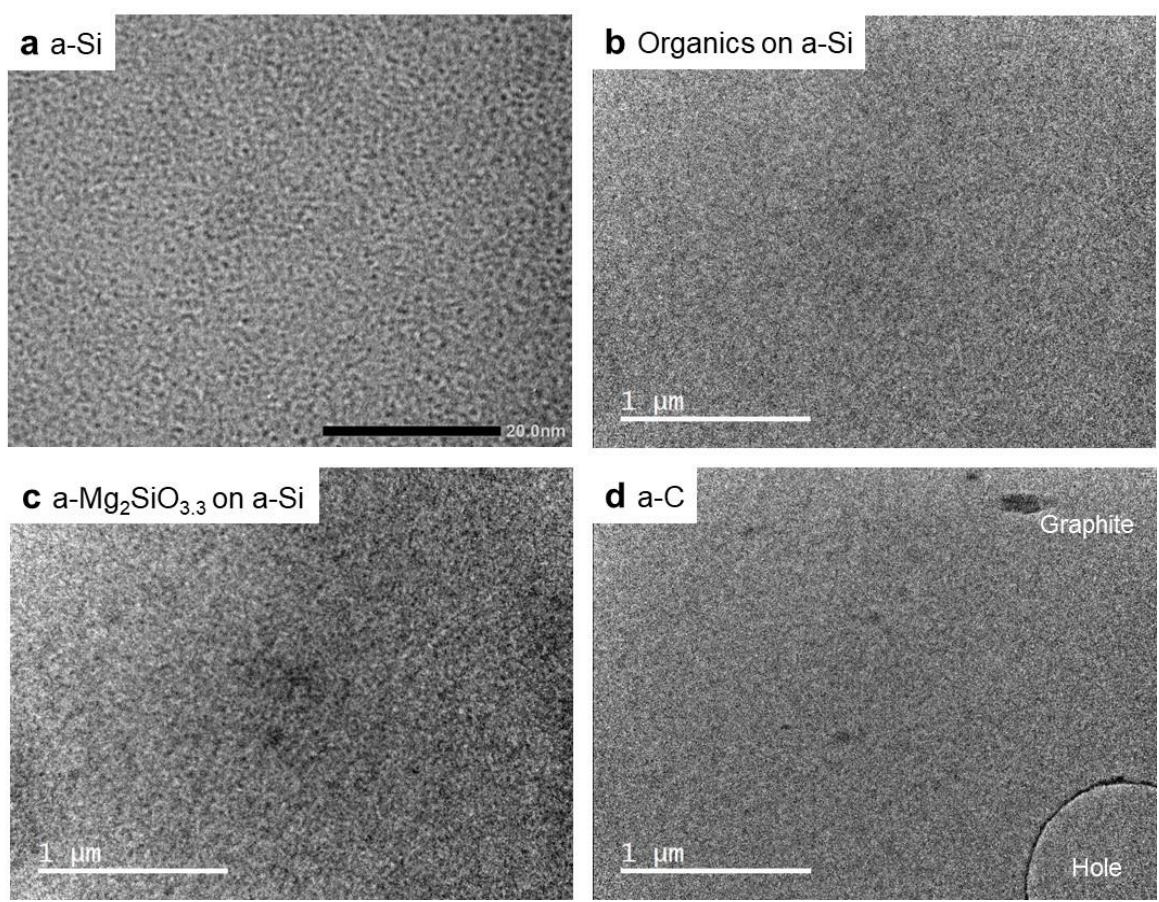

Figure 1. TEM images of amorphous thin films used as substrates: a) a-Si, b) organics formed on a-Si, c) a-$Mg_2SiO_{3.3}$ sputtered on a-Si, and d) a-C.

### 2.3. Experimental method

#### 2.3.1. Deposition of H$_2$O on the organics, a-Mg$_2$SiO$_{3.3}$, and a-C

H$_2$O was deposited onto each of the three substrates—organics, a-Mg$_2$SiO$_{3.3}$, and a-C with a deposition rate of ~1 nm minute$^{-1}$ at temperatures between 83 and 145 K. We observed the entire deposition process *in situ* using UHV-TEM.

#### 2.3.2. Crystallization of amorphous ices

Greenberg (1998) proposed an interstellar-dust model based on analyses of dust from Halley's comet, as shown in Table 1. However, the amounts of CO$_2$ and CO are smaller than those found in newer comets such as comet Hale-Bopp (e.g., Bockelée-Morvan et al. 2004). We thus modified the ice composition of Greenberg's model to H$_2$O:CO$_2$:CO=10:3:3 (volume ratio). If we assume that interstellar dust is spherical with a radius of 100 nm, the radius of the silicate core and the thicknesses of the other materials are calculated as shown in Table 1. Therefore, the deposition of a 19-nm-thick layer of H$_2$O onto the 26-nm-thick organics is ideal for mimicking astrophysical phenomena. However, thicker substrates disturb the TEM observation of thinner ice because the TEM contrast of amorphous material has a positive relationship to the atomic numbers and thicknesses of the samples (scattering contrast). We therefore use thinner substrates (10–20-nm thick) and relatively thicker ices (20–40-nm thick), as shown in Table 2. The contrast of crystalline samples is much stronger than that of the amorphous sample owing to the diffraction contrast; therefore, the detection of crystals on the amorphous substrate is easier than that of amorphous material on the amorphous substrate (Kouchi et al. 2021).

Table 1

Interstellar-dust models of Greenberg (1998) and the present study

|  |  | Silicate | Organics | Ices | | |
|---|---|---|---|---|---|---|
|  |  |  |  | $H_2O$ | $CO_2$ | CO |
| Greenberg | Mass fraction | 0.26 | 0.23 | 0.31 | 0.02 | 0.03 |
| model | Volume ratio | 0.1 | 0.27 | 0.36 | 0.02 | 0.04 |
| Present | Volume ratio | 0.1 | 0.27 | 0.36 | 0.11[a] | 0.10[a] |
| study | h[b] (nm) | 47.3 | 26.0 | 18.6 | 4.4 | 3.7 |

Note.

[a] Modified components

[b] Radius of silicate core or thickness of organics and ices when the radius of an interstellar grain is assumed to be 100 nm

Table 2

Experimental conditions for the crystallization of amorphous ices

| Amorphous substrates | | | Amorphous ices deposited | | | Crystallization | |
|---|---|---|---|---|---|---|---|
| Composition | Deposition temperature (K) | Thickness (nm) | Composition | Deposition temperature (K) | Thickness (nm) | Observation temperature (K) | Figure |
| $CO_2$ | 10 | 10 | CO | 10 | 20 | 21 | 4a |
| $H_2O$ | 10 | 20 | CO | 10 | 20 | 24.5 | 4b |
| $H_2O:CO_2$ = 5:1 | 10 | 20 | CO | 10 | 20 | 24 | 4c |
| $H_2O$ | 10 | 20 | $CO_2$ | 10 | 20 | 50, 60 | 5 |
| $H_2O$ | 10 | 20 | $CO:CO_2$ =1:1 | 10 | 30 | 13–70 | 6 |
| Organics | – | > 10 | $H_2O$ | 83 | 40 | 140 | 7a |
| $Mg_2SiO_{3.3}$ | 300 | 10 | $H_2O$ | 83 | 40 | 140 | 7b |
| C | – | 20 | $H_2O$ | 83 | 40 | 140 | 7c |
| Si | – | 5 | $H_2O:CO_2$ = 5:1 | 10 | 20 | 10–150 | 8 |

The procedures for ice deposition and crystallization are as follows. For experiments using a-$H_2O$ substrate, a 20-nm-thick a-$H_2O$ layer was firstly deposited at 10 K onto the a-Si substrate with a deposition rate of ~8 nm minute$^{-1}$; then, a 20-nm-thick a-CO layer was deposited at a rate of ~1 nm minute$^{-1}$. The same deposition rate was applied for a-$CO_2$ and a mixture of CO and $CO_2$ (a-CO:$CO_2$). When a-$H_2O$ was deposited onto the organics, a-$Mg_2SiO_{3.3}$, and a-C, the deposition temperature and deposition rate were set to 83 K and ~8 nm minute$^{-1}$, respectively. After deposition, the sample was heated at 10–30 K minute$^{-1}$ to the desired temperatures for a-CO, a-CO$_2$, and a-CO:CO$_2$; for a-H$_2$O, heating took place at ~10 K minute$^{-1}$.

### 2.3.3. TEM observation

We observed the deposition and heating process throughout using UHV-TEM. To avoid electron-beam damage to the samples, a low-dose technique (Tachibana et al. 2017) was applied using an 80-kV accelerating voltage, a very weak electron-beam intensity (~6 × 10$^{-3}$ electrons Å$^{-2}$ at the sample position), and a low-magnification observation (×25,000). When the electron-beam intensity had this value, we could not see any image on the fluorescent screen; we thus observed TEM images and electron-diffraction patterns using a CCD camera. All electron-diffraction patterns were collected in the 700-nm-diameter circular region of the central part of the TEM images. A brief explanation for the contrast of the TEM images of amorphous and crystalline ices was given by Kouchi et al. (2021).

### 2.3.4. Infrared spectroscopy

To measure the crystallization time scales of a-CO$_2$, we additionally used IR spectroscopy because the grain sizes of the CO$_2$ crystals at 35 and 40 K were too small to detect the termination of crystallization with TEM using the low-dose technique. The details of the experimental setup and procedure were described by Tsuge et al. (2020). Briefly, CO$_2$ gas was deposited onto a Si single-crystal substrate at 8 K by background deposition at a rate of $2 \times 10^{13}$ molecules cm$^{-2}$ s$^{-1}$ for 90 minutes. The prepared a-CO$_2$ was warmed at a rate of 4 K minute$^{-1}$ to the desired temperature for crystallization, and the crystallization was monitored *in situ* by FT-IR in the transmission mode with a resolution of 4 cm$^{-1}$.

## 3. Results

### *3.1. Deposition of H$_2$O onto various substrates*

Figure 2 shows the TEM images of ice I$_c$ deposited onto the organic, a-Mg$_2$SiO$_{3.3}$, and a-C substrates taken after saturation of the crystal number at temperatures between 145 and 130 K. At temperatures below 125 K, uniform films of a-H$_2$O were formed for all substrates. The number of crystals clearly decreases with increasing temperature. The number densities of ice I$_c$ crystals on a-Mg$_2$SiO$_{3.3}$ and a-C take almost the same values, but those on the organics are smaller at 145 and 140 K.

Based on the number density of crystals as measured via visual counting, we calculated the mean diffusion distances of H$_2$O molecules following Kouchi et al. (2020) (Fig. 3). According to Smith (1995), the mean surface diffusion distance, *X,* is expressed by $X = a(\nu n_0/F)^{1/2} \exp(-E_{sd}/2RT)$, where *a* is the hopping distance, $\nu$ the frequency factor, n$_0$ the number of adsorption sites, *F* the deposition flux, $E_{sd}$ the activation energy of surface diffusion, R the gas constant, and *T* the temperature. When *F* is constant, this behavior appears as a straight line with a negative slope of $-E_{sd}/2R$ on the Arrhenius plot of ln*X* vs. 1/*T* in Fig. 3. Consequently, from the slopes of these plots, we found the activation energies ($E_\mathrm{sd}$) of the surface diffusion of $H_2O$ molecules on the organics, a-$Mg_2SiO_{3.3}$, and a-C to be 2,830 ± 200 K, 1,940 ± 150 K, and 2,050 ± 500 K, respectively. These values are smaller than the activation energy of the surface diffusion of $H_2O$ on a-$H_2O$, which was estimated by Berland et al. (1995) to be 3,575 K; however, they are almost the same as (or larger than) that estimated by Zondlo et al. (1997) (2,115 K). In addition, we confirmed that ice $I_c$ also shows a lower wettability on these substrates in the case of deposition (see the 145-K images in Fig. 2). The $E_\mathrm{sd}$ values on a-$Mg_2SiO_{3.3}$ and a-C are 0.33 and 0.35 of the corresponding desorption energies on $H_2O$ (i.e., 5,964 K and 5,928 K), respectively, as measured by Potapov et al. (2018).

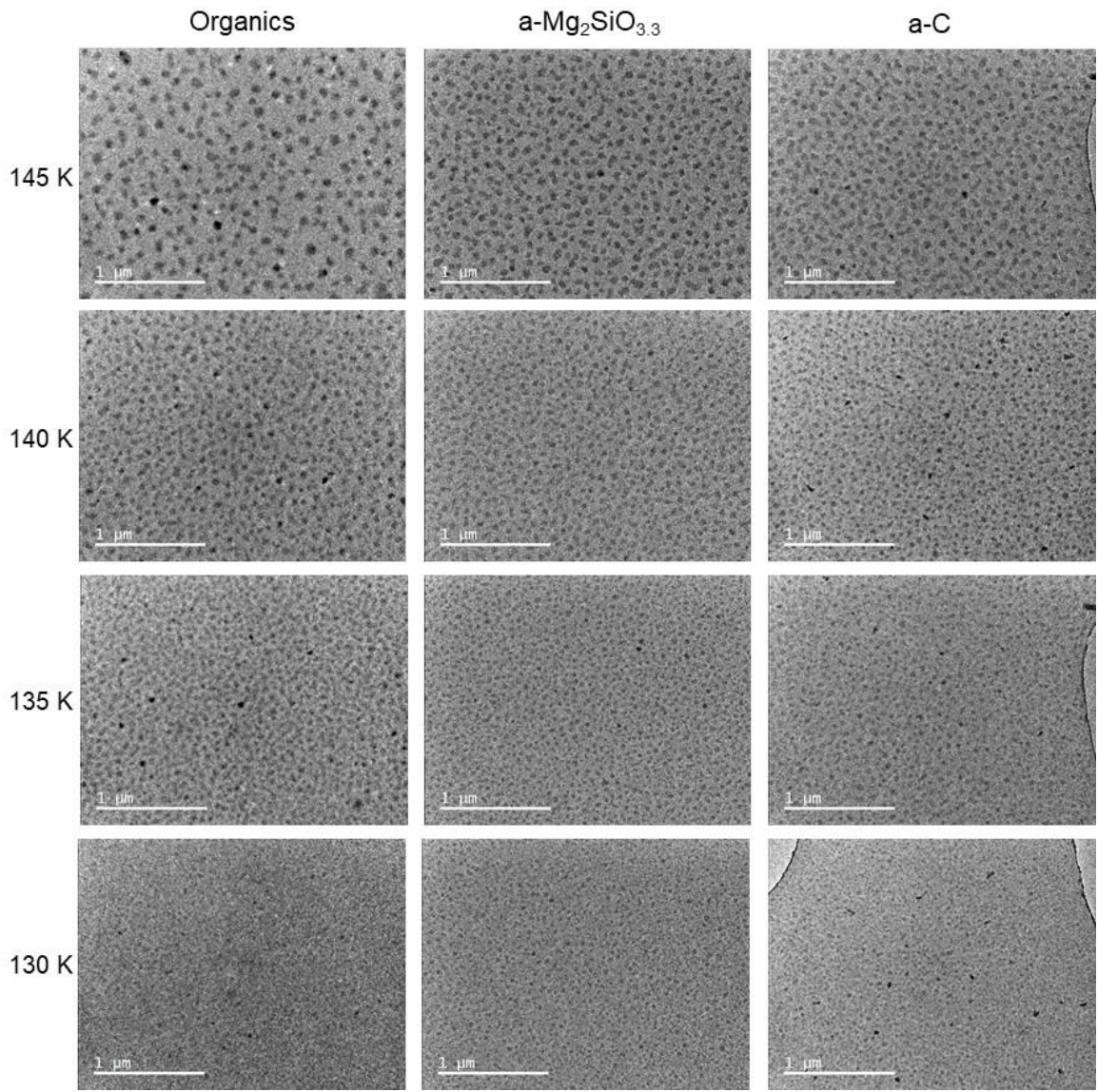

Figure 2. TEM images taken just after saturation of the ice $I_c$ crystal number densities during the deposition of $H_2O$ onto the organics, a-$Mg_2SiO_{3.3}$, and a-C at 145–130 K.

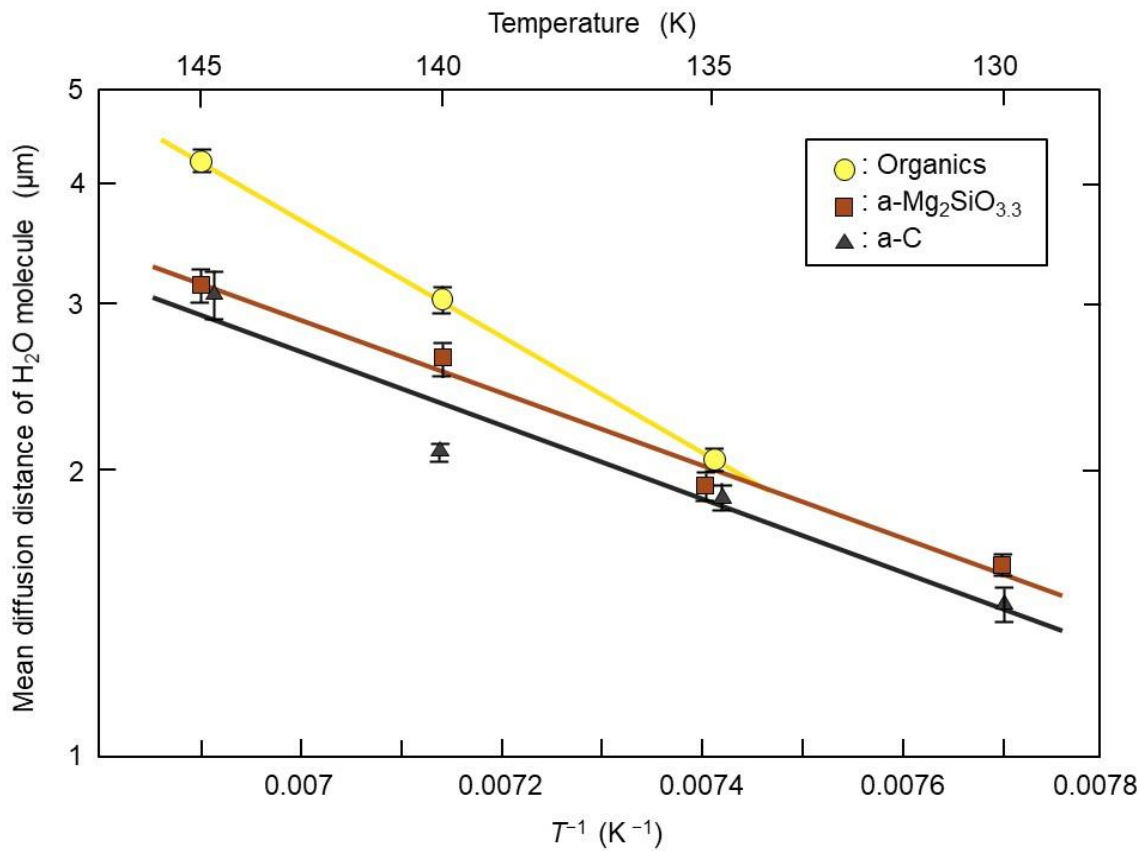

Figure 3. Plots of the mean diffusion distance of H$_2$O molecules on various substrates vs. inverse temperature.

### 3.2. Crystallization of amorphous ices

#### 3.2.1. Crystallization of a-CO

Figure 4 shows the crystallization of a-CO on the a-CO$_2$, a-H$_2$O, and a-H$_2$O:CO$_2$ substrates. The TEM images of CO at 10 K are uniform and electron diffraction shows halo patterns, suggesting that these samples are amorphous CO (a-CO) (Fig. 4a). TEM images of CO at 21 K suggest the occurrence of crystals because of the strong diffraction contrast (Kouchi et al. 2021); the electron-diffraction pattern confirms that crystalline CO (α-CO) was formed. In the image taken after 2 minutes at 21 K, the crystal size was uniform and ~50 nm in size. After 8 minutes, coarsening of the crystals was observed. TEM images clearly show that α-CO did not grow as a uniform film but as three-dimensional islands.

Similar phenomena were observed in the cases of crystallization of a-CO on the a-$H_2O$ (Kouchi et al. 2021) and a-$H_2O$:$CO_2$ = 5:1 substrates; however, in these cases, the formation of three-dimensional islands was remarkable (Figs. 4b, c). After 0.2 minutes at 24.5 K (Fig. 4b), crystals were uniformly around 50 nm in size; after three minutes, only specific crystals grew to ~200 nm in size, with others remaining the same size or becoming smaller. We usually call this phenomenon Ostwald ripening (e.g., Lifshitz & Slyozhov 1961; Bhakta & Ruckenstein 1995).

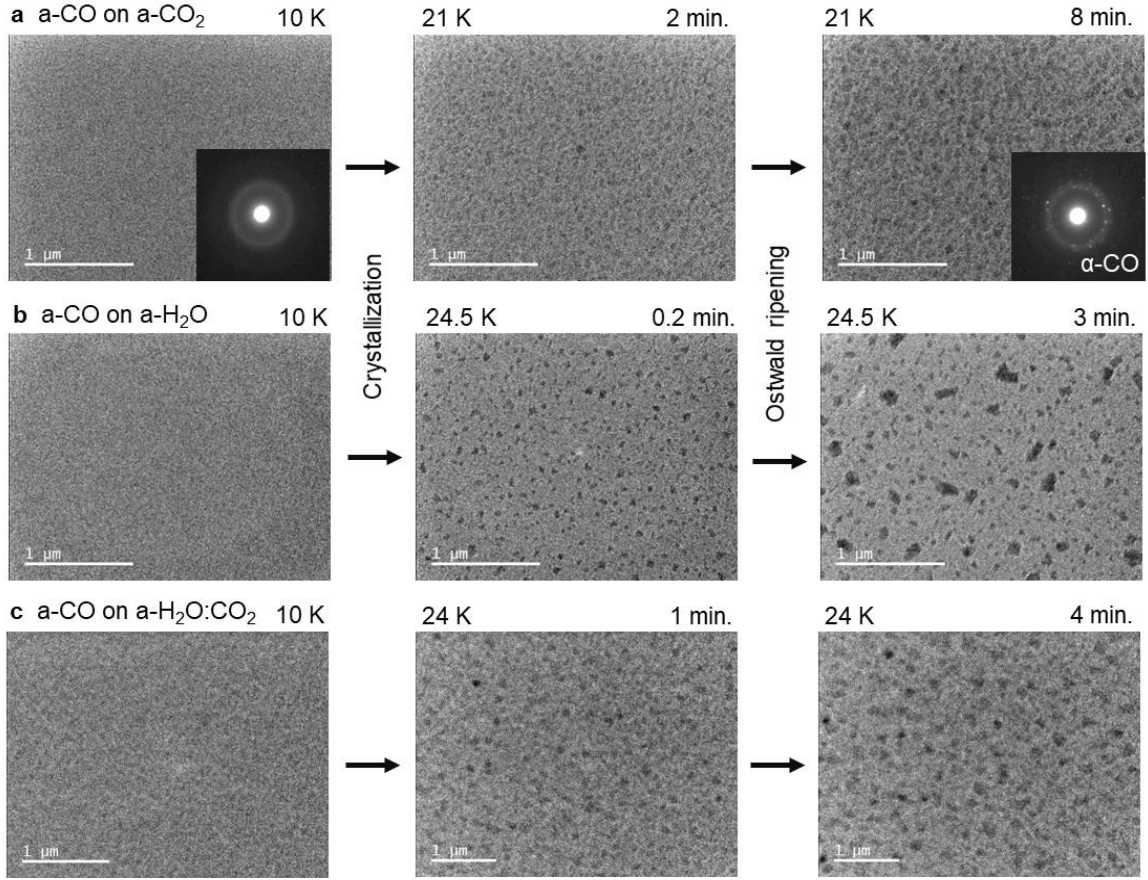

Figure 4. TEM observation of the crystallization of a-CO on (a) a-CO$_2$, (b) a-H$_2$O, and (c) a-H$_2$O:CO$_2$ = 5:1. The corresponding electron-diffraction patterns are shown in some images.

### 3.2.2. Crystallization of a-CO$_2$

Figure 5 shows the crystallization of a-CO$_2$ on the a-H$_2$O substrate at 50 K and 60 K. In the TEM images of the 50-K samples, almost no change can be observed, but the electron-diffraction pattern clearly shows the formation of CO$_2$ I crystals. The size of the crystals ($d$) could be calculated from the full-width-at-half-maximum of the diffraction peak in terms of the $2\theta$ angle, $B(2\theta)$ using

$$d = 0.85\lambda/[B(2\theta)\cos\theta], \tag{1}$$

where λ is the wavelength of an 80-kV electron (Ida et al. 2003). The size of the $CO_2$ I crystals after 10 minutes at 50 K is 9.3 nm. At 60 K, the size increased to ~50 nm, showing that the coarsening proceeded in $CO_2$ I; however, the $CO_2$ crystals formed did not show a three-dimensional island structure like α-CO, but rather formed an almost uniform film.

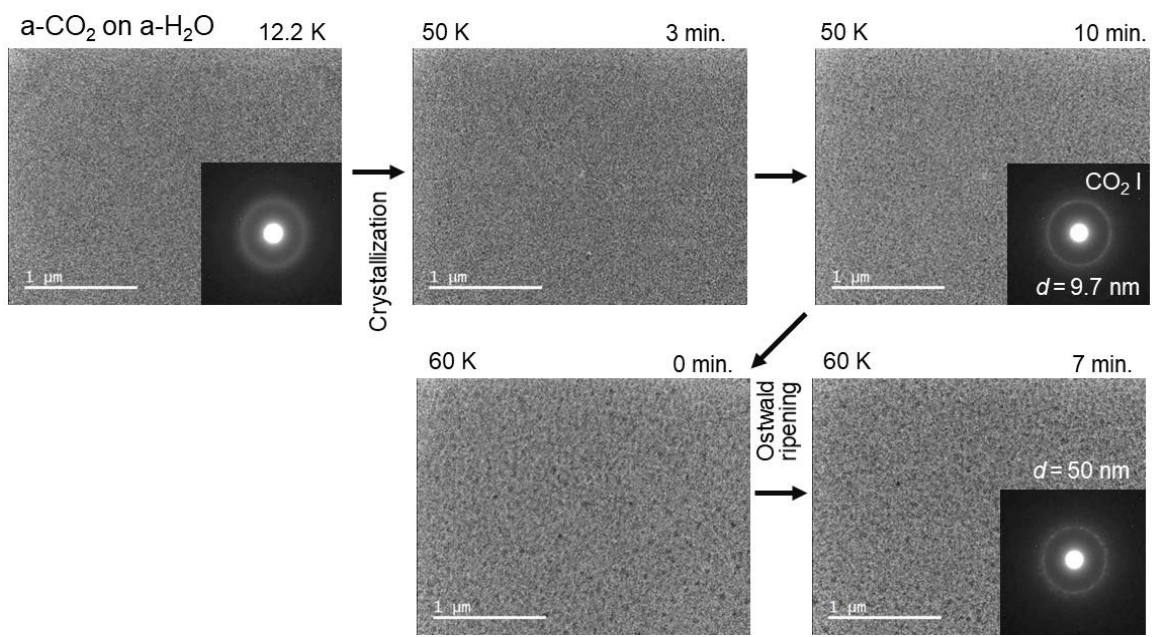

Figure 5. TEM observation of the crystallization of a-$CO_2$ on a-$H_2O$ substrate at 50 K and 60 K.

### 3.2.3. Crystallization of a-CO:$CO_2$

The changes of the TEM images and corresponding electron-diffraction patterns during heating of a-CO:$CO_2$=1:1 are shown in Fig. 6. At first glance, TEM images appear not to be changed from 12.8 K to 70 K; however, electron-diffraction patterns clearly show the occurrence of crystalline phases: α-CO with $d > 3.6$ nm at 33 K, $CO_2$ I with $d = 4.1$ nm at 50 K, and $CO_2$ I with $d > 10$ nm at 70 K. We note that the crystallization temperature of a-CO (between 25 K and 33 K) is higher than that of pure a-CO (<21 K, Fig. 4a; ~20 K, Kouchi 1990); this may be due to the slow bulk diffusion of CO in the $CO:CO_2$ matrix.

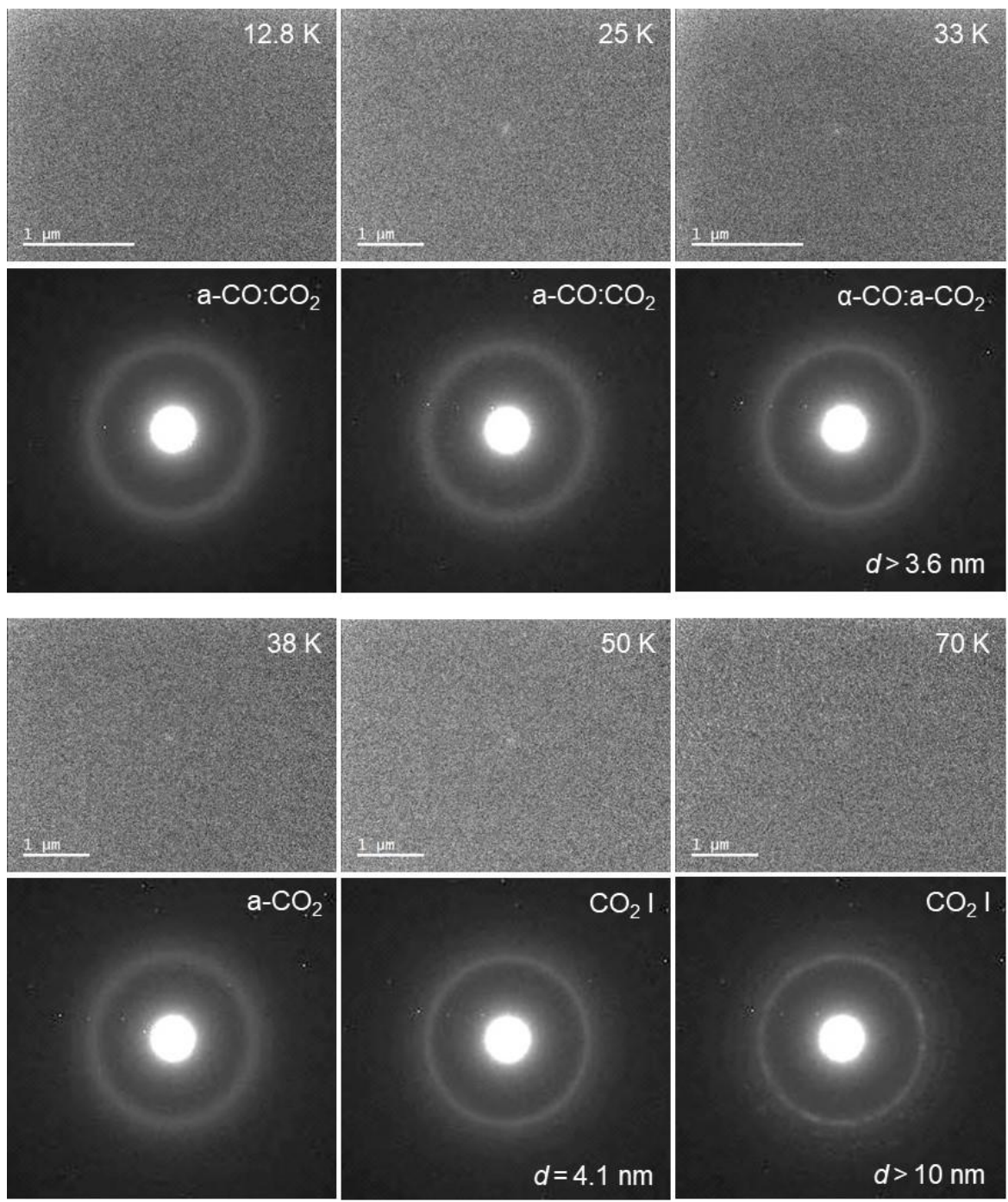

Figure 6. TEM images and corresponding electron-diffraction patterns of the crystallization of a-CO:CO$_2$ on a-H$_2$O substrate from 12.8 K to 70 K. Note that sublimation of α-CO occurs between 33 and 38 K.

### 3.2.4. Crystallization of a-$H_2O$

The crystallization processes of a-$H_2O$ on the organics, a-$Mg_2SiO_{3.3}$, and a-C substrates are shown in Fig. 7. After one minute at 140 K, small three-dimensional crystalline islands of ~50 nm were formed on all substrates. Then, these islands grew to sizes of 100–200 nm via Ostwald ripening. This is consistent with the observation of the crystallization of a-$H_2O$ on a-C (Jenniskens & Blake 1996: Fig. 4). Furthermore, the fact that there is no difference among the three kinds of substrates suggests that the interfacial energies between ice $I_c$ and the three substrates have similar values.

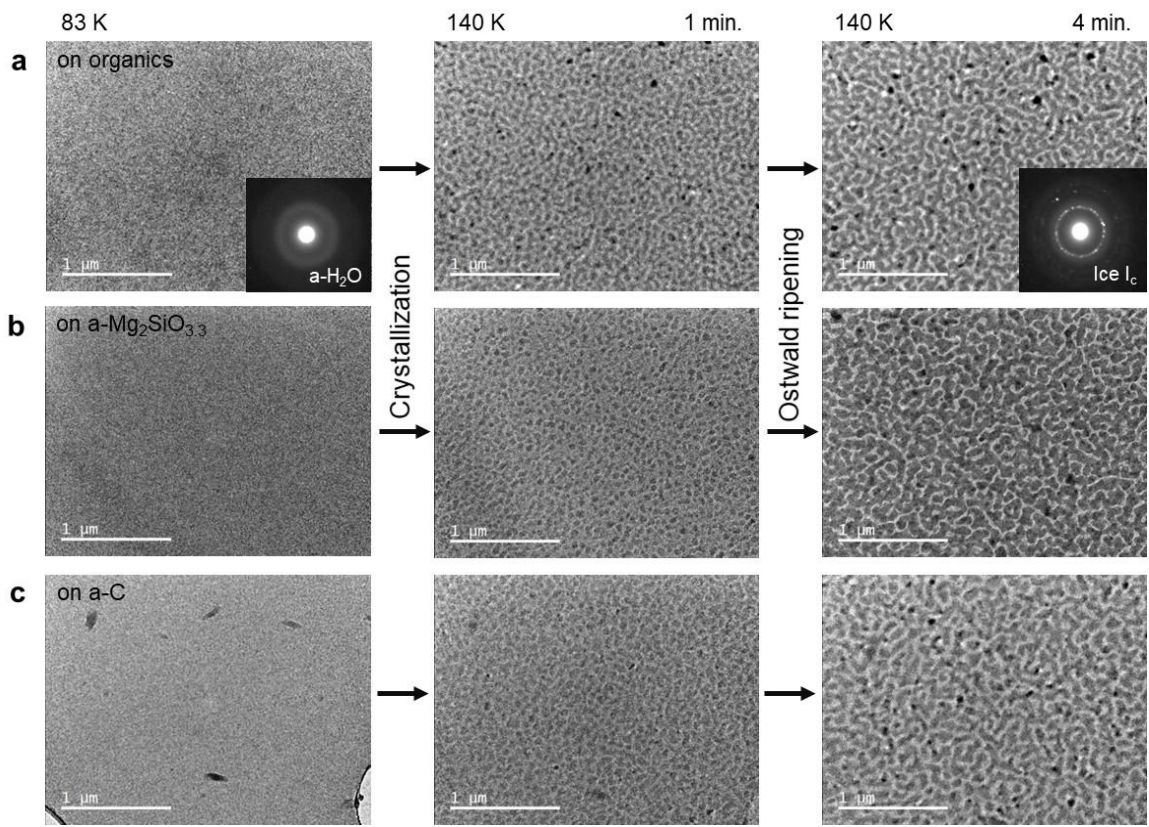

Figure 7. TEM observation of the crystallization of a-$H_2O$ at 140 K on various substrates: a) Organics, b) a-$Mg_2SiO_{3.3}$, and c) a-C.

*3.2.5. Crystallization of a-$H_2O$:$CO_2$*

The changes in the TEM images and corresponding electron-diffraction patterns during heating of a-$H_2O$:$CO_2$ = 5:1 are shown in Fig. 8. At 10 K, a-$H_2O$:$CO_2$ was a uniform mixture; at around 60 K, $CO_2$ I started to form as nano crystals, with some crystalline nuclei growing to 50 nm at 70 K, as shown in Fig. 8b. At around 80 K, $CO_2$ I sublimated and a-$H_2O$ remained at temperatures above 80 K (Fig. 8c). The remaining $CO_2$ may have been included in a-$H_2O$ as an impurity, although TEM could not reveal the $CO_2$ content. At 140 K, crystallization of a-$H_2O$ started to form 50-nm ice $I_c$ (Fig. 8d). The present results for the crystallization temperature of a-$CO_2$ and the sublimation temperature of $CO_2$ I are consistent with previous studies using infrared spectroscopy (e.g., Sandford & Allamandola 1990a; Hodyss et al. 2008; Öberg et al. 2009b).

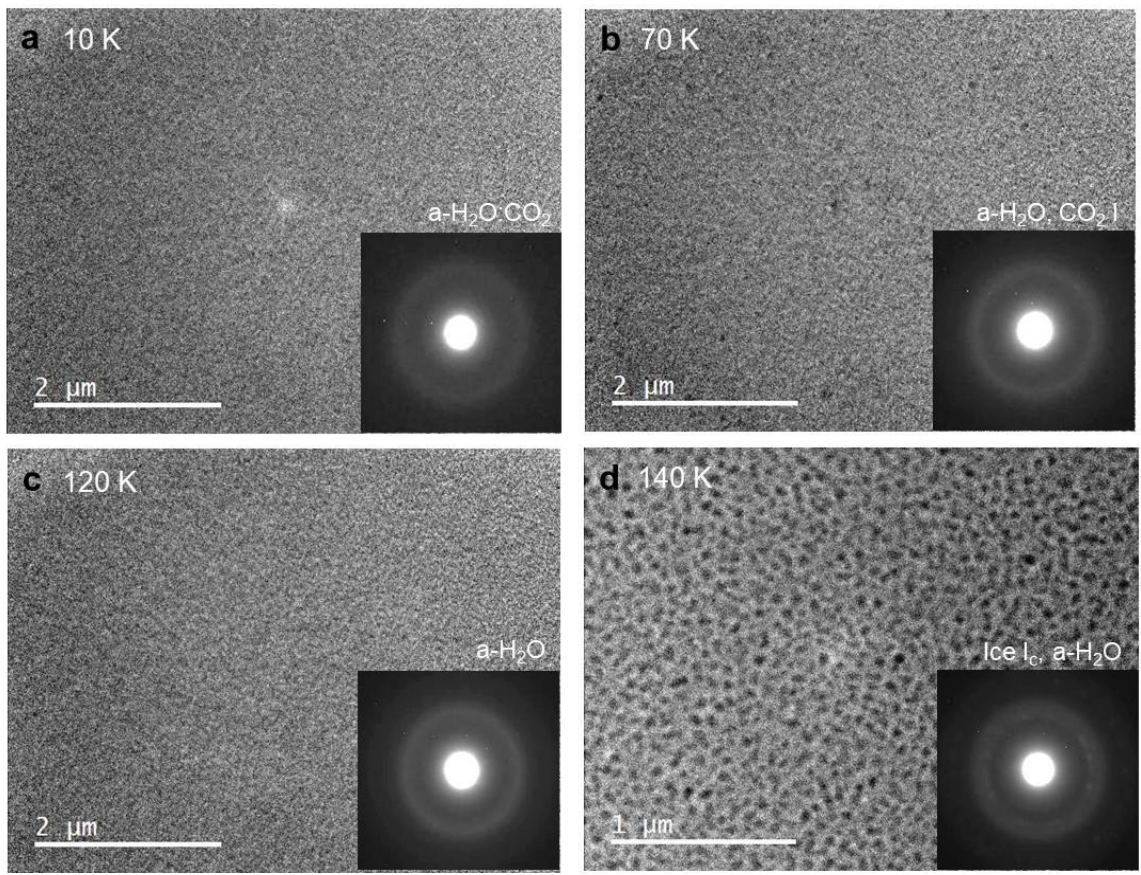

Figure 8. TEM images and corresponding electron-diffraction patterns of the crystallization of a-$H_2O$:$CO_2$ on a-Si under heating from 10 K to 140 K at a rate of ~1 K minute$^{-1}$.

### 3.2.6. Summary of the crystallization observation

Observations of crystallization are schematically summarized in Fig. 9. When a-CO crystallized on the a-$H_2O$ and a-$CO_2$ substrates, three-dimensional islands were formed; then, Ostwald ripening proceeded. When a-$H_2O$ crystallized on the organics, a-$Mg_2SiO_{3.3}$, and a-C substrates, similar processes were observed in all cases. The sizes of these crystals became ~200 nm, which are the almost the same scale as icy grains in space, strongly suggesting the occurrence of one crystal of α-CO on the icy grain of a molecular cloud and one ice I crystal on the refractory grain in a proto-planetary disk. On the other hand, when a-$CO_2$ crystallized on the a-$H_2O$ substrate, filmy nano-crystals were formed; then, coarsening occurred. The wetting of crystalline ices on the substrates varies depending on the combination of ice and substrate. α-CO on a-$H_2O$ shows the lowest wettability, and ice $I_c$ on the organic, a-$Mg_2SiO_{3.3}$, and a-C substrates shows a poor wettability. Meanwhile, $CO_2$ I on a-$H_2O$ shows something between a high wettability and perfect wetting. These wettability and crystalline sizes greatly affect the morphology of icy grains in both molecular clouds and proto-planetary disks, as will be discussed later.

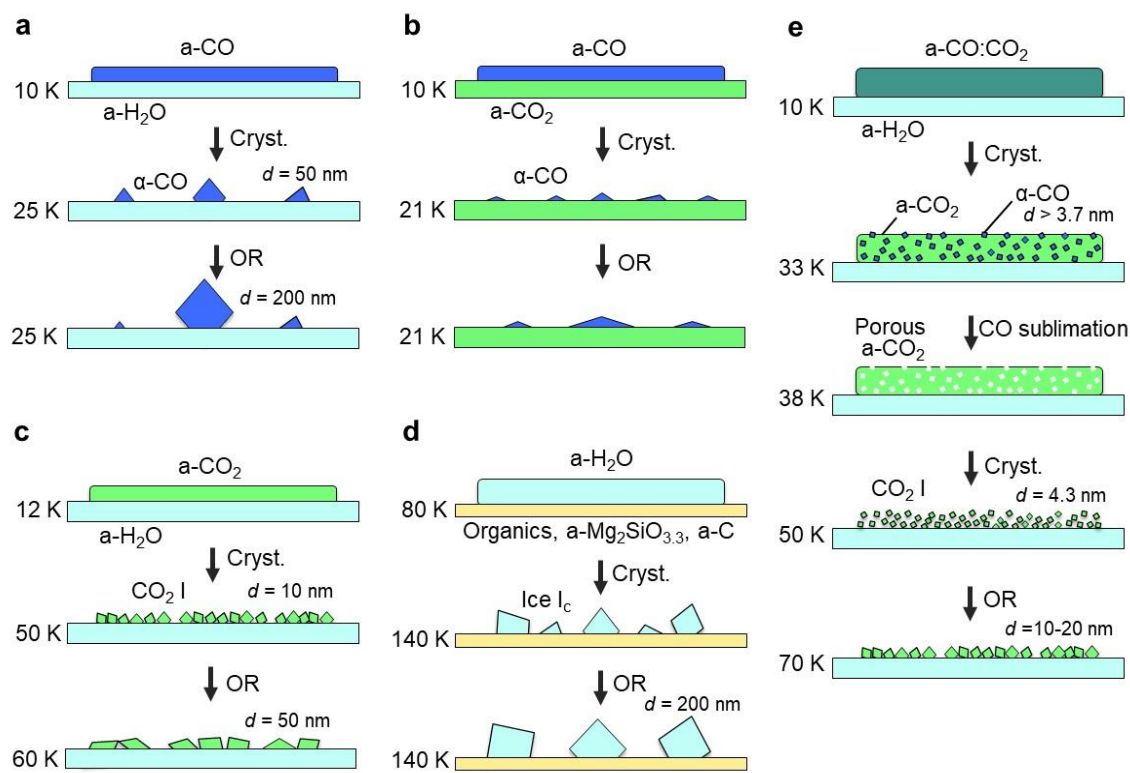

Figure 9. Schematic illustration of the crystallization processes observed

in this study. Respective colors show the chemical compositions of ices and substrates; Blue: CO, light blue: $H_2O$, green: $CO_2$, dark green: $CO:CO_2$, and yellow: organics, $Mg_2SiO_{3.3}$, C. Cryst: crystallization, OR: Ostwald ripening.

*3.2.7. Timescale of crystallization*

We measured the crystallization timescale of a-$CO_2$; in general, detecting the onset of crystallization is not so difficult using TEM, but detecting the end point is rather difficult. Therefore, our measurement may have a large error of about one order of magnitude. In the case of CO, measurement was not so difficult using TEM because the size of the crystals is relatively large (Kouchi et al. 2021). Meanwhile, in the case of $CO_2$, measurement was very difficult (especially at lower temperatures) because determining the end point of the crystallization was complicated by the crystals' nm-size (Fig. 5, 6). Therefore, we measured the change in the IR spectra of a-$CO_2$ at 35 and 40 K, as shown in Fig. 10, and determined the timescale of crystallization, as shown in Fig. 11. At temperatures above 50 K, on the other hand, measurement by the IR spectrum becomes very difficult because it takes a long time to heat the substrate from 10 K to 50 K; i.e., crystallization will be completed during heating processes. The crystallization timescale, $t_{cryst}$, of a-$CO_2$ is expressed by $t_{cryst} = A \exp(E/kT)$, where $k$ is Boltzmann's constant, $T$ represents the temperature, and $A$ and $E$ are constants determined from the fitting: $A = 5.3 \times 10^{-3}$ s and $E/k = 540$ K. In Fig.

11, $t_{\text{cryst}}$ for a-CO (Kouchi et al. 2021) and a-$H_2O$ (Kouchi et al. 1994) are also shown. By extrapolating these data to lower temperatures, we estimated the timescale of crystallization to be $10^3$ years at 10 K for a-CO and $10^5$ years at 16 K for a-$CO_2$.

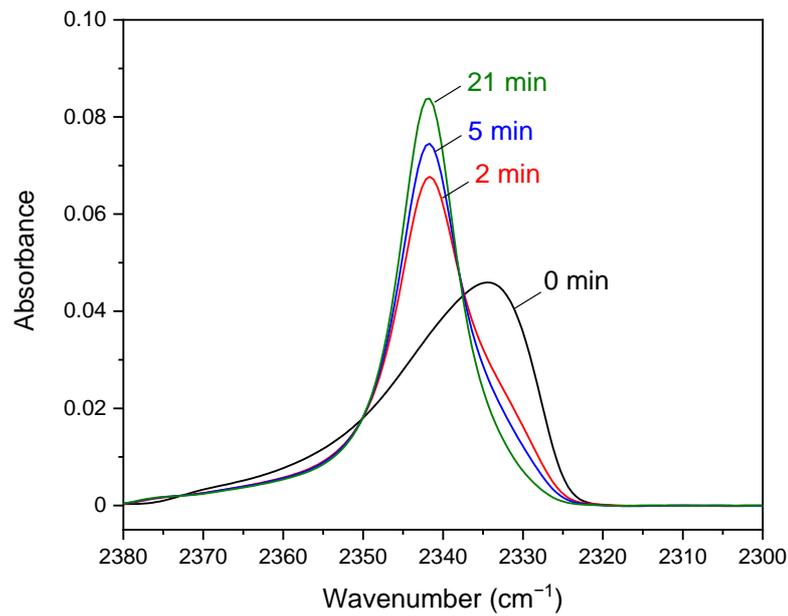

Figure 10. Change in the IR spectra during crystallization of a-$CO_2$ at 40 K. The peak positions for a-$CO_2$ and $CO_2$ I are at 2,333 and 2,342 cm$^{-1}$, respectively (Tsuge et al. 2020).

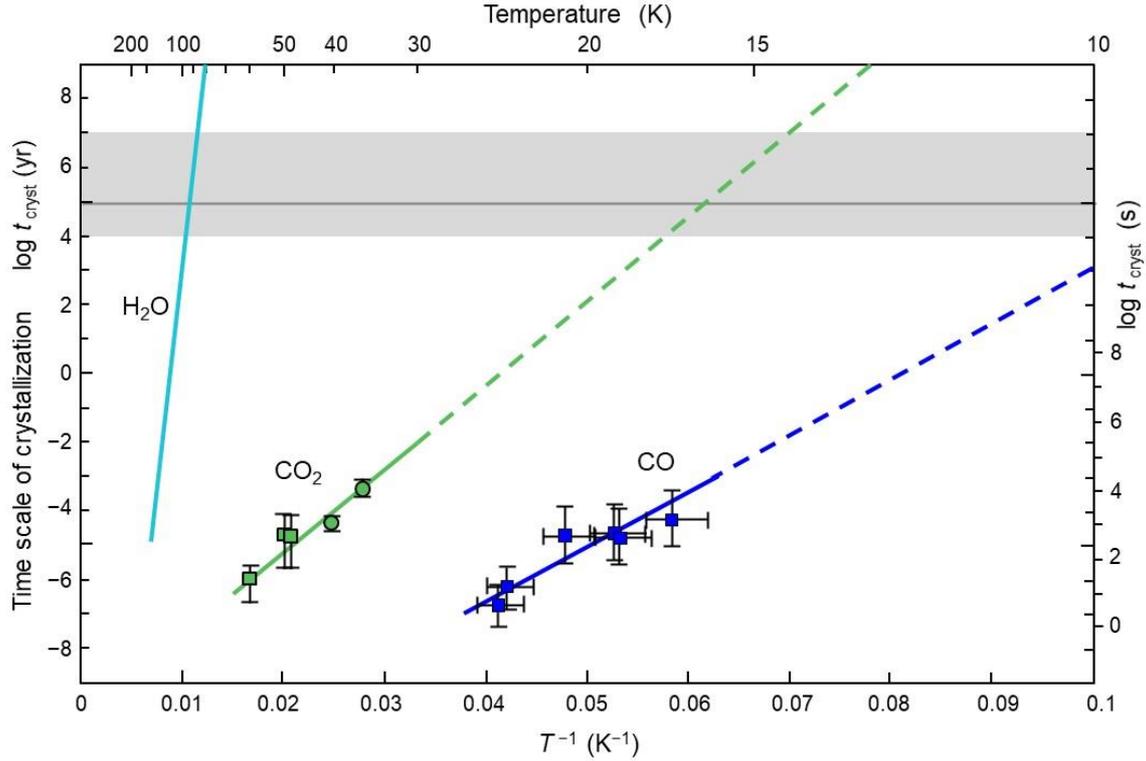

Figure 11. The timescales of 100% crystallization for various ices. Green line: a-$CO_2$ measured in this study. $CO_2$ at 35 and 40 K were measured using IR spectroscopy, while others were measured by TEM observation. Blue line, a-CO (Kouchi et al. 2021); light blue line: a-$H_2O$ (Kouchi et al. 1994). The broken lines show extrapolations to lower temperatures. The gray shaded area is the relevant timescale for discussions of molecular clouds and proto-planetary disks and the gray horizontal line indicates $10^5$ years.

4. Discussion

*4.1. Surface-diffusion coefficients*

To discuss the crystal morphology on the grain surface, information concerning the surface-diffusion coefficient is essential. Figure 12 shows the surface-diffusion coefficients, $D_s$, of CO on a-$H_2O$, $CO_2$ on a-$H_2O$, and $H_2O$ on organics, a-$Mg_2SiO_{3.3}$, and a-C. $D_s$ is expressed as:

$$D_s = D_0 \exp(-E_{sd}/kT), \qquad (2)$$

where $D_0$ is the pre-exponential factor, $E_{sd}$ is the activation energy of the surface diffusion, $k$ is Boltzmann's constant, and $T$ is the temperature. Kouchi et al. (2020) measured the $E_{sd}$ of CO and $CO_2$ on a-$H_2O$ to be 350 and 1,500 K, respectively. Thus, we can calculate $D_s$ if $D_0$ is assumed to be $a^2\nu$ where $a$ is the hopping distance and $\nu$ is the hopping frequency ($a$ = 0.3 nm and $\nu$ = $10^{12}$ s$^{-1}$). Note that our discussion is not affected by the uncertainty of $\nu$, because the error of $\nu$ is smaller than a few orders of magnitude. There has been no direct measurement of the surface self-diffusion coefficients of CO and $CO_2$; therefore, we assume the activation energies of surface diffusion to be 0.3 times those of desorption (Sandford & Allamandola 1988, 1990b), as shown by the broken lines. The $D_s$ of CO on the (111) face of α-CO (see Fig. 15) is calculated based on the theoretical estimation of the adsorption energy for that face (see Section 5.1), as shown by a blue broken line. The $D_s$ of $H_2O$ on organics, a-$Mg_2SiO_{3.3}$, and a-C could also be calculated using the measured activation energies (shown by yellow, brown, and black solid lines in Fig. 12). The $D_s$ of $H_2O$ on the (0001) face of ice $I_h$ (Kiefer & Hole 1977) is also shown by a light blue broken line. Note that the surface structure of the (0001) face of ice $I_h$ is the same as that of the (111) face of ice $I_c$ (see Fig. 15). From these results, we conclude that CO molecules diffuse almost freely on a-$H_2O$, even in 10-K molecular clouds, while $CO_2$ molecules do not. $CO_2$ molecules on a-$H_2O$ and $H_2O$ molecules on organics start to diffuse at temperatures above ~32 and ~63 K, respectively.

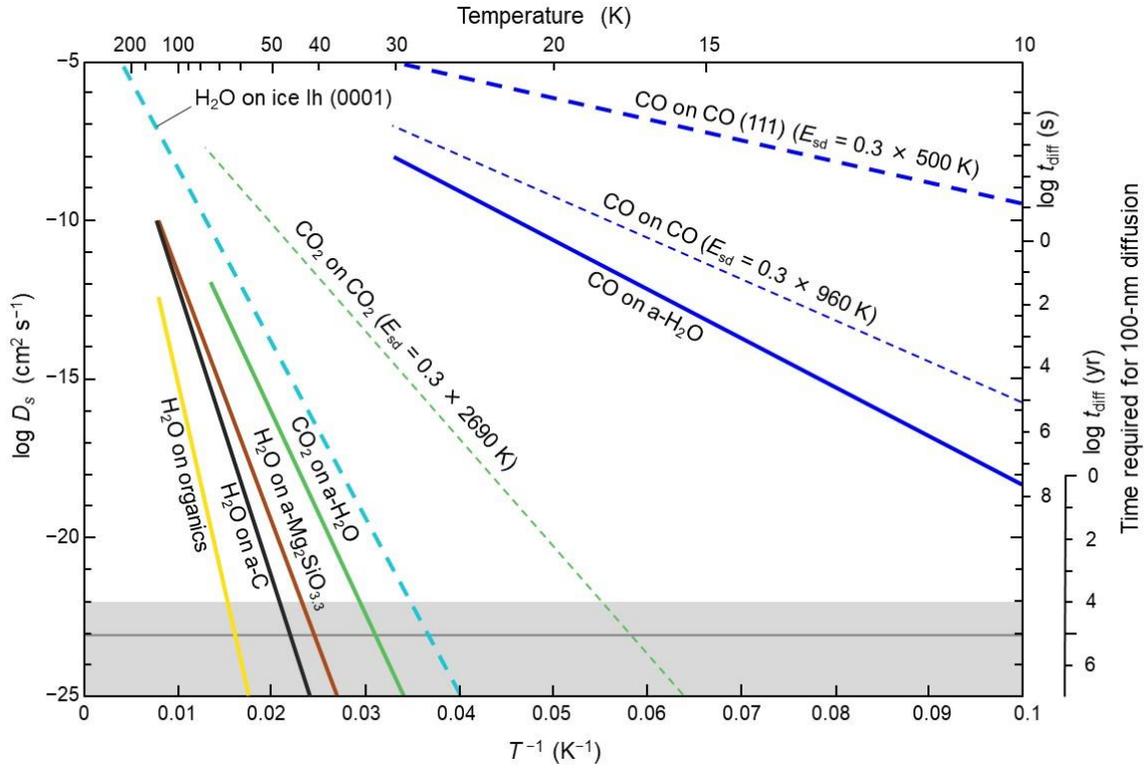

Figure 12. Surface-diffusion coefficients. Solid yellow, green, and blue lines show the surface-diffusion coefficients of $H_2O$ on organics (this study), $CO_2$ on a-$H_2O$ (Kouchi et al. 2020), and CO on a-$H_2O$ (Kouchi et al. 2020), respectively. The thick broken light-blue and blue lines show the surface-diffusion coefficients of $H_2O$ on the (0001) face of ice $I_h$ (Kiefer & Hole 1977) and CO on the (111) face of α-CO, respectively. The latter is calculated by assuming that the activation energy of the surface diffusion is 0.3 times the adsorption energy of 500 K (see Table 3). Thin broken green and blue lines show the surface-diffusion coefficients of $CO_2$ on $CO_2$ and CO on CO, respectively, as calculated under the assumption

that the activation energies of surface diffusion are 0.3 times the desorption energies obtained by Sandford & Allamandola (1988, 1990b). On the right ordinate, the time required for 100-nm diffusion on the substrates is also shown. The gray shaded area is the relevant time scale for discussing the evolution of molecular clouds and proto-planetary disks, and the gray horizontal line indicates $10^5$ years.

4.2. Crystallinity of ices formed in molecular clouds

Kouchi et al. (1994) present the condition for the formation of amorphous ice during deposition:

$$F > F_c \equiv D_s / a^4, \qquad (3)$$

where $F$ is the flux of molecule, $F_c$ is the critical flux above which amorphous ice is deposited, and $a$ is the hopping distance. Figure 13 shows the $F_c$ values for a-$H_2O$ on organics, a-$CO_2$ on a-$H_2O$, and a-CO on a-$H_2O$. Following Kouchi et al. (1994), the fluxes of O atoms ($F_O$) and CO molecules ($F_{CO}$) are described by

$$F_O = n_{H_2} f_O (kT/2\pi\mu_O m_H)^{1/2}, \qquad (4)$$

and

$$F_{CO} = n_{H_2} f_{CO} (kT/2\pi\mu_{CO} m_H)^{1/2}, \qquad (5)$$

respectively, where $n_{H_2}$ is the number density of $H_2$ molecules, $f$ indicates fractional abundances, $\mu$ indicates molecular weight, and $m_H$ is the mass of a hydrogen atom. If we assume that $n_{H_2} = 10^4$ cm$^{-3}$, $f_O = 10^{-4}$, $f_{CO} = 4 \times 10^{-5}$, then $T = 10$ K, $F_O = 2.7 \times 10^3$ cm$^{-2}$ s$^{-1}$ and $F_{CO} = 1.1 \times 10^3$ cm$^{-2}$ s$^{-1}$. It is clear that crystalline

CO (α–CO) is formed by the deposition of CO in 10-K molecular clouds, because $F_{CO}$ is about ten orders of magnitude smaller than the critical flux of CO (Kouchi et al. 2021). Because $H_2O$ and $CO_2$ molecules are formed by surface-atomic reactions (as discussed in the next section), $F_{H_2O}$ and $F_{CO_2}$ could be assumed to be equal to $F_O$ and $F_{0.5CO}$, respectively. We conclude that a-$H_2O$ and a-$CO_2$ are formed in 10-K molecular clouds because the effective fluxes of $H_2O$ and $CO_2$ are much larger than the critical fluxes.

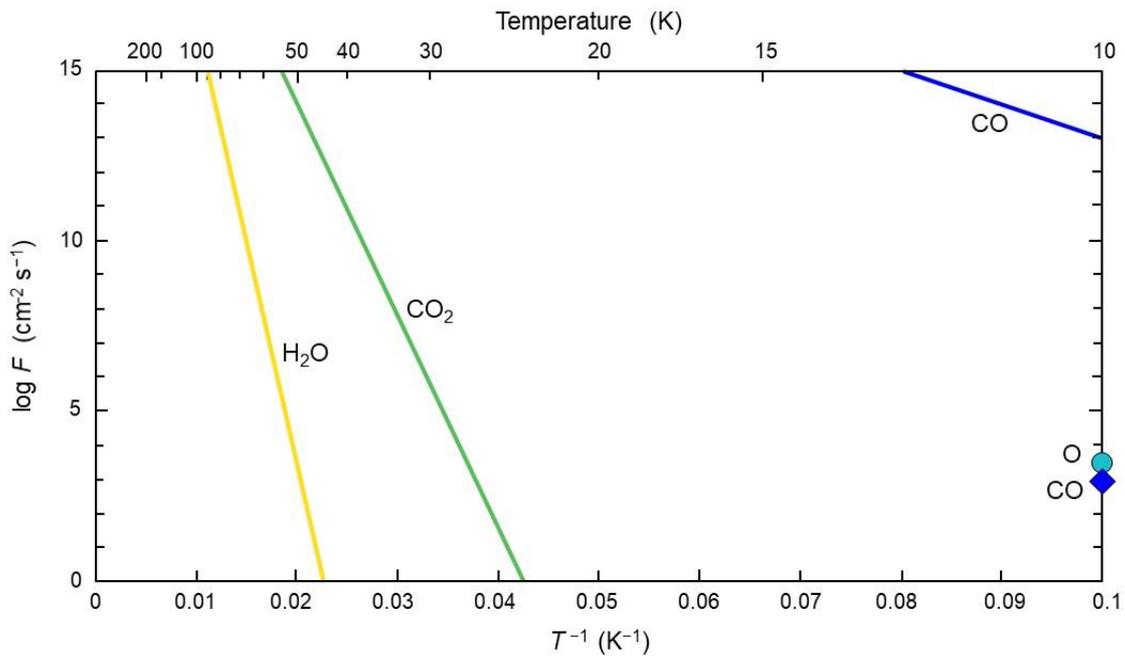

Figure 13. The critical fluxes of $H_2O$, $CO_2$, and CO, above which amorphous ices are formed. The critical fluxes of $H_2O$ and $CO_2$ are calculated using the surface-diffusion coefficients of the respective molecules in Fig. 12 ($CO_2$ on a-$H_2O$ and $H_2O$ on organics), and that of CO on a-$H_2O$ is from Kouchi et al. (2021). The fluxes of O and CO in the 10-K molecular clouds are shown by a light-blue circle and a blue diamond, respectively.

*4.3. Formation process of ices in molecular clouds*

A spherical (or ellipsoidal) layered structure has been implicitly assumed for icy grains in molecular clouds in most previous works (e.g., Ehrenfreund et al. 1998; Pontoppidan et al. 2008). Only Pontoppidan et al. (2003) suggested that CO is crystalline α-CO and that α-CO takes the form of small, irregularly shaped clumps on top of the $H_2O$ ice mantle. Therefore, it is desirable to find other evidence to check this assumption. Furthermore, there has been no discussion of the morphology of $CO_2$ ice. We will investigate the formation processes of ices on the refractory grains in molecular clouds based on the calculation results for the chemical evolution of ices (e.g., Garrod & Pauly 2011; Furuya et al. 2015; Ruaud et al. 2016; Kouchi et al. 2020). The compositional changes shown by ices from the inner part to outer are as follows: pure $H_2O$ to $H_2O:CO:CO_2$ (basic cloud model of Garrod & Pauly 2011), $H_2O:CO_2$ to $H_2O:CO:CO_2$ (three-phase model of Ruaud et al. 2016), and pure $H_2O$, through $H_2O:CO_2$, to $CO:H_2O$ (Furuya et al. 2015; Kouchi et al. 2020). Although the details differ among these studies, one can extract a general tendency; inner layers have $H_2O$-rich composition, intermediate layers $H_2O:CO_2$, and outer layers $H_2O:CO:CO_2$.

At first, the $H_2O$-rich layer was grown by the accumulation of $H_2O$ molecules formed by the following surface reactions (Miyauchi et al. 2008; Ioppolo et al. 2008; Dulieu et al. 2010; Oba et al. 2012; Hama & Watanabe 2013):

$$O + H \rightarrow OH + \Delta E \ (429 \ kJ \ mol^{-1}), \ OH + H \rightarrow H_2O + \Delta E \ (499 \ kJ \ mol^{-1}), \quad (6)$$

$$OH + H_2 \rightarrow H_2O + H + \Delta E \text{ (62 kJ mol}^{-1}\text{), and} \tag{7}$$

$$O_2 + H \rightarrow HO_2 + \Delta E \text{ (203 kJ mol}^{-1}\text{), } HO_2 + H \rightarrow H_2O_2 + \Delta E \text{ (354 kJ mol}^{-1}\text{),} \tag{8}$$

$$\text{and } H_2O_2 + H \rightarrow H_2O + OH + \Delta E \text{ (285 kJ mol}^{-1}\text{).}$$

Because $H_2O$ molecules were formed from O atoms as above, the flux of O atoms in 10-K molecular clouds determined the crystallinity of $H_2O$ ice. The flux of O atoms is much larger than the critical flux of $H_2O$, above which a-$H_2O$ formed (Fig. 13), indicating that the $H_2O$ ice formed was a-$H_2O$. Additionally, the fact that UV-irradiation promotes amorphization (Kouchi & Kuroda 1990) also supports this conclusion. Therefore, a-$H_2O$ uniformly covered the refractory grains (Fig. 14a). Oba et al. (2009) suggested that the ice formed by surface reactions has a compact structure compared to vapor-deposited a-$H_2O$, because the spectral features due to dangling OH bonds were not observed in the IR spectrum.

The composition of the next layer is expected to be $H_2O:CO_2$ = 3:2–5:1. Formation of the $CO_2$ molecule itself proceeded mainly according to the following grain-surface reaction (Yu et al. 2001; Oba et al. 2010; Garrod & Pauly 2011):

$$CO + OH \rightarrow CO_2 + H + \Delta E \text{ (96 kJ mol}^{-1}\text{).} \tag{9}$$

The fluxes of the CO and O atoms in 10-K molecular clouds are considerably larger than the critical flux of a-$CO_2$ formation (Fig. 13), indicating that the solid $CO_2$ formed was initially amorphous if it had no effect on the heat of the reaction. Simultaneous formation of $H_2O$ and $CO_2$ led to the growth of the mixed layer. It has been implicitly assumed that this $H_2O:CO_2$ mixed ice was amorphous (Pontoppidan et al. 2008); however, the following facts suggest the occurrence of $CO_2$ crystals, i.e., $CO_2$ I:

a. Even if a-$CO_2$ was formed, a-$CO_2$ crystallized within $10^5$ years at 16 K (Fig. 11); UV-irradiation promotes crystallization of a-$CO_2$ (Tsuge et al. 2020); and the heat of the $H_2O$, $CO_2$, and $H_2$ (434 kJ mol$^{-1}$)-formation reactions may promote crystallization. If we assume that the whole layer was formed over $10^6$ years, there might be a small amount of a-$CO_2$ on the uppermost layer, as shown in Fig. 14.

b. Direct formation of crystalline $CO_2$ was possible because the $CO_2$ molecule formed by reaction (9) possesses a sufficient energy to diffuse momentarily on a-$H_2O$, leading to crystallization (Tsuge et al. 2020).

c. Because $CO_2$ molecules did not diffuse on a-$H_2O$ at around 10 K (as shown in Fig. 12) and because the wetting of $CO_2$ I to a-$H_2O$ is good (Fig. 5, Fig. 9c), the $CO_2$ crystals formed should be of nanometer size and were embedded in a-$H_2O$, as shown in Fig. 14b.

Pure crystalline $CO_2$ ice as well as $H_2O$:$CO_2$ ice mixtures shows a characteristic double-peak profile in the 15 μm bending mode (e.g., Ehrenfreund et al. 1996; Gerakines et al. 1999; Bergin et al. 2005). Such a profile has not been observed in prestellar cores (Bergin et al. 2005; Whittet et al. 2007), suggesting the absence of $CO_2$ crystals in dense starless cores. On the other hand, Escribano et al. (2013) suggested the occurrence of crystalline $CO_2$ in dense clouds based on the absence of the 2,328 cm$^{-1}$ band towards Elias 16, where the band traces pure a-$CO_2$. For further confirmation, Mie scattering by ellipsoidal or irregularly shaped grains, which will change the band profile, needs to be evaluated. Given the estimated timescale of $CO_2$ crystallization (i.e., $10^5$ yr at 16 K with a sharp temperature dependence, Fig. 11), it is unclear whether the crystallization of the $CO_2$ ice is fully completed in a prestellar stage. The present crystallization process will be applicable to the crystalline $CO_2$ observed in the line of sight towards embedded young stellar objects (Pontoppidan et al. 2018). In those sources, we anticipate that the crystallized $CO_2$ will form nano-crystals $CO_2$ I in a-$H_2O$ as shown in Fig. 14. The infrared spectral variation of crystalline $CO_2$ at various measurement conditions has yet to be investigated, particularly in the 15 μm bending mode region. Because the present IR setup uses an HgCdTe(MCT) detector that covers 2–13 μm, the spectrum in the 15 μm region could not be measured. Future studies incorporating TEM observations, IR spectroscopy, and comparison with astronomical spectra for various crystalline ice samples are awaited. Moreover, high quality measurements with the upcoming James Webb Space Telescope *(*JWST) mission may shed more light on the crystallinity of $CO_2$ ice.

The composition of the outer layer is $H_2O$:CO:$CO_2$, becoming CO-rich with time. Pontoppidan et al. (2003) suggested the occurrence of α-CO from an IR observation of embedded young low-mass stars and performed a detailed analysis of the spectra. The experimental results obtained by Kouchi et al. (2021) strongly support their suggestion because (1) the flux of CO in the 10-K molecular clouds is smaller than the critical flux for the deposition of a-CO (Fig. 13); (2) the timescale of crystallization of a-CO at 10 K is only $10^3$ years (Fig. 11); and (3) UV-irradiation promotes crystallization of a-CO. Pontoppidan et al. (2003) also suggested that α-CO are small, irregularly shaped clumps on top of the $H_2O$ ice mantle; however, our present study shows that solid CO was deposited as one single crystal, α-CO (Figs. 14c, d), because CO molecules diffused almost freely on a-$H_2O$ due to the very large diffusion coefficient (Fig. 12) and accumulated on the largest crystal. In other words, Ostwald ripening proceeded very efficiently. Because the wetting of α-CO against a-$H_2O$ is bad (as shown in Figs. 4b, c and Fig. 9a), α-CO is not embedded in a-$H_2O$ nor forms a uniform layer. Thus, icy grain would have a morphology where an α-CO crystal is attached on the a-$H_2O$ (Fig. 14d). The above discussion has been somewhat simplified; we do not deny the occurrence of small amounts of CO and $CO_2$ in a-$H_2O$ as impurities and very small clusters.

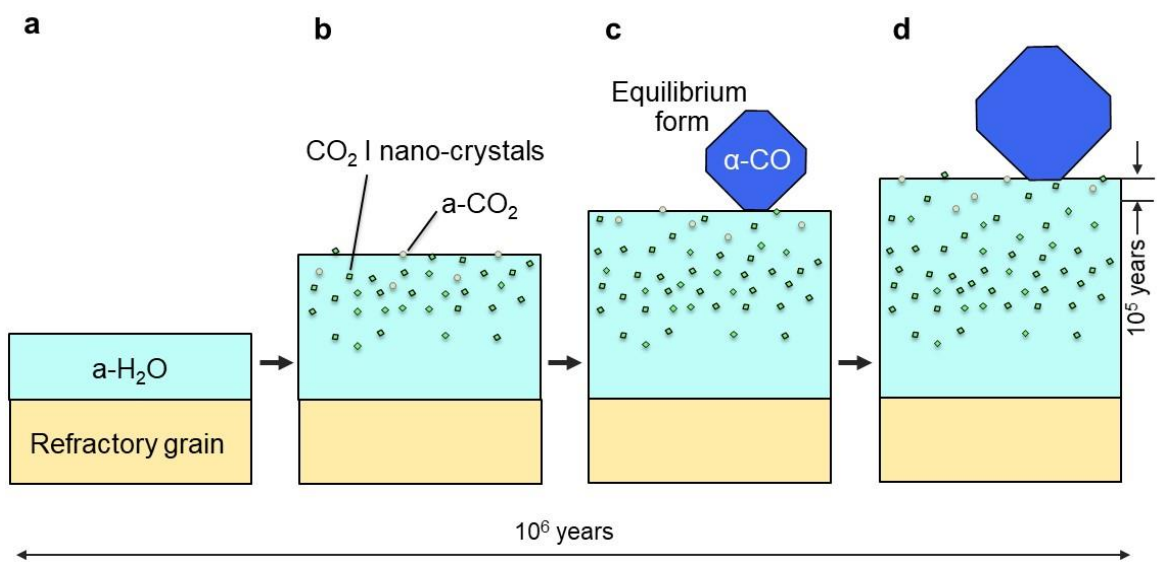

Figure 14. Formation and evolution of ices on the refractory grains in molecular clouds.

*4.4. Factors affecting crystal morphology*

First, we will discuss equilibrium forms on ice crystals to consider the morphology of icy grains in space. The equilibrium form of a crystal is defined as the crystal form at which thermodynamic equilibrium is attained. In this case, the Gibbs free energy of the system (i.e., the summation of volume energy and surface energy) should be minimized. The equilibrium form of the crystal becomes a polyhedron with flat crystalline faces. As shown in Fig. 15, the equilibrium forms of ice $I_h$, ice $I_c$, and face-centered-cubic (FCC) crystal are a hexagonal prism (Krastanow 1943), a regular octahedron (Takahashi 1982), and a truncated regular octahedron (Toschev 1973), respectively. When the self-diffusion coefficients of respective molecules on the corresponding crystals are sufficiently large, equilibrium forms should be realized. Since the surface self-diffusion coefficients of $H_2O$ on ice $I_h$, $CO_2$ on $CO_2$, and CO on CO are sufficiently large (as shown in Fig. 12), the respective crystals become their equilibrium forms. Therefore, it is reasonable to assume that thin-layered ice crystals are unstable because the surface energy of thin-layered ice crystals is a few times larger than that of a small sphere of the same volume. We note that we do not know whether $CO_2$ I nano-crystals in a-$H_2O$ take on their equilibrium form or not due to the lack of diffusion data concerning $CO_2$ in a-$H_2O$.

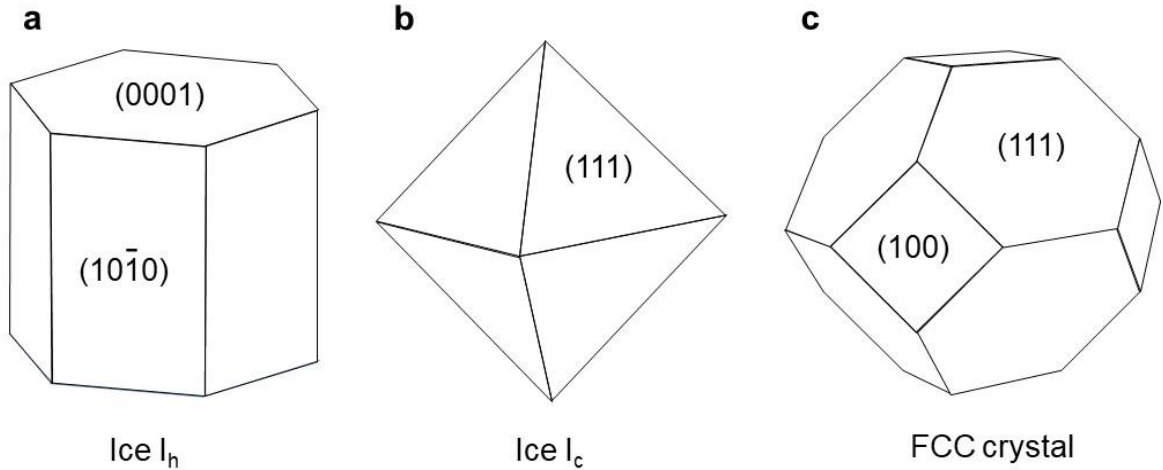

Figure 15. Equilibrium forms of crystals: a) ice $I_h$ (Krastanow 1943), b) ice $I_c$ (Takahashi 1982), and c) FCC crystal (Toschev 1973).

The second factor affecting icy-grain morphology is the wetting of ices against substrates; typically, the degree of wetting is determined by Young's equation,

$$\gamma_{sub} = \gamma_{sub\text{-}ice} + \gamma_{ice} \cos\theta, \qquad (10)$$

where $\gamma_{sub}$ and $\gamma_{ice}$ are the surface energies of the substrate and the ice crystal, respectively, $\gamma_{sub\text{-}ice}$ is the interfacial energy between the substrate and the ice crystal, and $\theta$ is the contact angle. When wetting is complete, $\theta = 0°$; the poorer the wetting, the larger $\theta$ becomes. However, for the wetting of α-CO on a-$H_2O$, the situation is not so simple because there is an adsorption layer between α-CO and a-$H_2O$ (Kouchi et al. 2021). We should consider the wetting of an adsorbed CO layer on a-$H_2O$, as well as that of α-CO on the adsorbed CO layer. Unfortunately, when coverage is larger than one monolayer, there are no quantitative data for CO and $CO_2$ adsorption on porous a-$H_2O$, or for $H_2O$ on organics, a-$Mg_2SiO_{3.3}$, and a-C. It should be noted that when coverage is smaller than one monolayer, complete wetting of CO on porous a-$H_2O$ was observed (He et al. 2016). This result looks somehow contradictory to the present study, but could be explained by considering surface energies and interfacial energy of the system, which change greatly depending on the coverage (Kouchi et al. 2021). Therefore, we will discuss the wetting of ices against various substrates qualitatively based on TEM observations. Poor wetting is observed in α-CO on a-$H_2O$ (Figs. 4b, c) and almost complete wetting is observed in $CO_2$ I on a-$H_2O$ (Fig. 5). When ice $I_c$ was formed on organics, a-$Mg_2SiO_{3.3}$, and a-C substrates, poor wetting was observed in all cases (Figs. 2, 7). Based on these results, we will discuss, in the next section, the morphologies of ices on various substrates and their evolution in proto-planetary disks.

The third factor is the number of crystals formed on the substrate, which is governed by the surface diffusions of the various molecules on the substrate. If we consider diffusion over $10^5$ years, CO on a-$H_2O$, $CO_2$ on a-$H_2O$, and $H_2O$ on organics could diffuse sufficiently to form one crystal on the corresponding substrates at temperatures above ~8, ~32, and ~63 K, respectively (Fig. 12). The crystallization temperatures of a-CO, a-$CO_2$, and a-$H_2O$ over $10^5$ years are ~9, ~16, and ~88 K, respectively (Fig. 11); this shows that the Ostwald ripening started simultaneously with the crystallization of a-CO and a-$H_2O$. In the case of $CO_2$, nano-crystals remained at temperatures between 10 and 32 K and Ostwald ripening started at temperatures above 32 K. Therefore, we conclude that each component (CO, $CO_2$ and $H_2O$) forms a single crystal with its equilibrium form (i.e., α-CO, $CO_2$ I, and ice I) on the grains at temperatures higher than ~9, ~32, and ~63 K, respectively.

*4.5. Evolution of icy grains in proto-planetary disks*

Figure 16 compares the evolution of icy grains in proto-planetary disks between the previous assumption and our new model. It has been assumed that all ices have layered structures regardless of composition (CO, $CO_2$, $H_2O$) or crystallinity (amorphous, crystalline). However, it is clear that this assumption is imprecise for the following reasons:

a. The flux of CO (Fig. 13) and the crystallization time scale of a-CO (Fig. 11) suggest the absence of a-CO in molecular clouds. Kouchi et al. (2021) showed that that UV-rays and high-energy electron irradiation never cause amorphization of α-CO.

b. Escribano et al. (2013) suggested an absence of a-$CO_2$ in molecular clouds, and this study supports their suggestion from the time scale of crystallization (Fig. 11). Tsuge et al. (2020) showed that UV irradiation never causes amorphization of crystalline $CO_2$.

c. A simple calculation shows that the total Gibbs free energy of the thin crystalline layer with large radius is larger than that of a small sphere of equivalent volume, because the surface energy of the former is larger than that of the latter.

Based on the experimental results of this study, we consequently propose a new model shown in Fig. 16. As already discussed above, grains in molecular clouds are composed of a silicate and organics core, a pure a-$H_2O$ layer, an a-$H_2O$ layer including nano-crystalline $CO_2$ I, and one α-CO crystal attached to the a-$H_2O$. When temperature exceeded the CO snow line, α-CO sublimated. Further increase in temperature above ~32 K caused efficient surface diffusion of $CO_2$ molecules on a-$H_2O$, resulting in the formation of one $CO_2$ I thin crystal on the a-$H_2O$. This picture differs greatly from the simple layered-ice model adopted in some previous studies of proto-planetary dust growth (e.g., Musiolik et al. 2016; Pinilla et al. 2017; Okuzumi & Tazaki 2019), which assume icy grains consisting of a refractory core, an inner $H_2O$ mantle, and an outer $CO_2$ mantle. When the temperature exceeded the $CO_2$-snow line, $CO_2$ I sublimated and the a-$H_2O$ grains remained ellipsoidal. A further increase in temperature above ~88 K caused efficient surface diffusion of $H_2O$ molecules on the organics, resulting in the formation of one ice I crystal. Note that this conclusion did not change when ice I was formed on other refractory substrates, a-silicate, or a-C, because the interfacial energies between ice I and the other materials have similar values. Gärtner et al. (2017) concluded from the scanning electron microscope observation of pure $H_2O$ ice particles of a few micrometers in size that the morphology of a crystalline ice particle is a truncated sphere and that collision experiments using ice spheres are good analogs for proto-planetary environments. However, our present results clearly show that crystalline ice particles are not truncated spheres but polyhedral crystals (Figs. 15a, b) attached to the refractory core (Fig. 16), and that collision experiments performed thus far using ice spheres are not good analogs for proto-planetary environments.

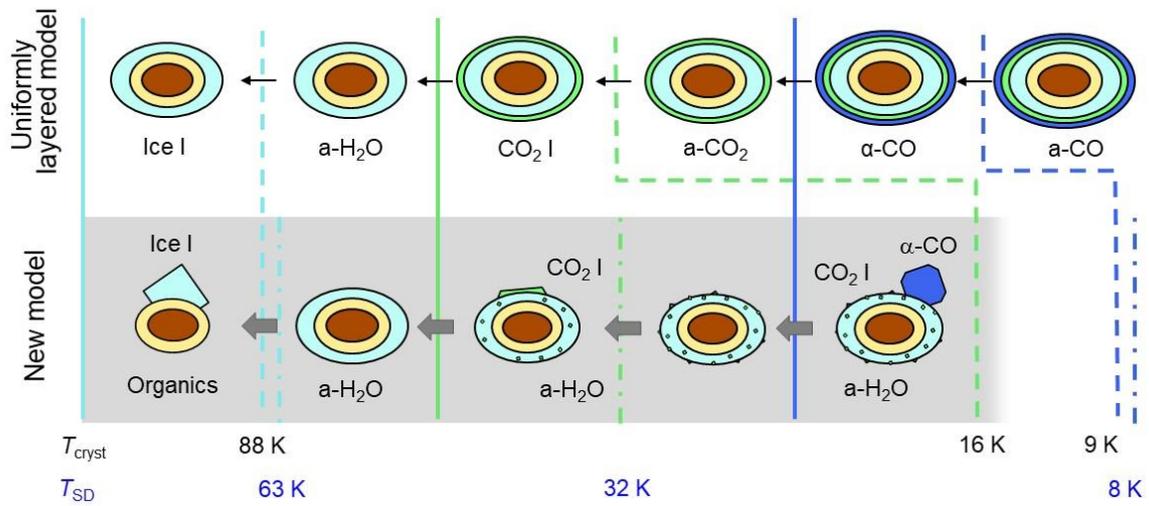

Figure 16. Schematic thermal-evolution models of icy grains in proto-planetary disks. Blue, green, and light blue show CO, $CO_2$, and $H_2O$, respectively. Vertical solid and broken lines show the snow and crystallization lines, respectively. The names of the outermost grain materials are shown. $T_{cryst}$ is the crystallization temperature observed over $10^5$ years. $T_{SD}$ (as shown by blue letters and vertical dashed-dotted-lines) are the temperatures at which CO, $CO_2$, and $H_2O$ could diffuse 100 nm on the substrates over $10^5$ years.

## 5. Astrochemical and astrophysical implications

### 5.1. Adsorption energies of atoms and CO on the α-CO surface

In general, the chemical evolution on grains depends on the concentration and mobility of atoms and molecules. These have been often discussed in terms of atomic and molecular adsorption and diffusion energies at the surface of a-$H_2O$. Here, we present our computational estimate of the adsorption energies of H, C,

N, and O atoms (as well as the CO molecule itself) onto the α-CO (111) surface model shown in Fig. 17.

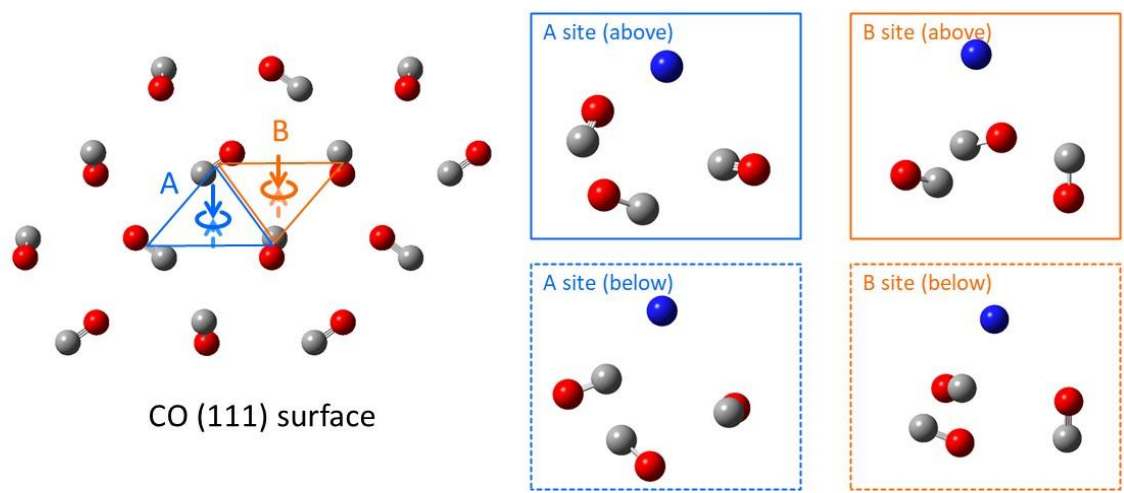

Figure 17. α-CO (111) surface model. Two different hollow sites (A and B) and adsorptions from above and below were taken into account. A blue sphere represents an adsorbate. The geometries of three CO molecules from the α-CO surface were fixed to the crystalline structure.

Quantum chemical calculations were performed using the Gaussian 09 program (Frisch et al. 2016) similarly to Shimonishi et al. (2018). We optimized the position of the adsorbing species (H, C, N, O, and CO) using the Møller–Plesset method (Møller & Plesset, 1934) with the cc-pVTZ basis set (MP2/cc-pVTZ level of theory), where stationary points on the potential energy surface were searched. For optimized structures, the total energy (i.e., single point energies) of the system, $E_{system}$, was calculated using the coupled-cluster method (Scuseria et al.

1988; CCSD(T)/cc-pVTZ level of theory), which correctly incorporates dispersion interactions between atoms and surface CO molecules. With the energies of the α-CO (111) surface model ($E_{adsorbent}$) and adsorbing species in vacuum ($E_{adsorbate}$), adsorption energies $E_{ads}$ were calculated as $E_{ads} = -[E_{system} - (E_{adsorbent} + E_{adsorbate}]$.

Table 3 summarizes the calculated adsorption energies on the α-CO (111) surface model. On the α-CO (111) surface, it should be noted that sites A and B have an appearance ratio of 1:3. From our results, H, N, O, and CO are weakly bound to the α-CO (111) hollow site due to physisorption; these adsorption energies are apparently smaller than those for the water-ice surface (Shimonishi et al. 2018), indicating that these atoms and molecules exhibit higher mobilities on the α-CO surface. On the other hand, the adsorption energy of the C atom at site B on the α-CO (111) surface was computed to be quite large (over 20,000 K); this is because the C atom reacts with a surface CO molecule to create a C=C=O molecule, as shown in Fig. 18; in other words, the C atom is strongly bound to the α-CO (111) surface due to chemisorption.

Consequently, our calculated results suggest that mechanisms of chemical evolution are to be changed on the α-CO surface due to the higher mobility of atoms and molecules and/or the formation of C=C=O molecules, which would be a source of organic molecules. Furthermore, it should be emphasized that different chemical evolutions proceed simultaneously on a-$H_2O$ and α-CO (Fig. 14) in 10-K molecular clouds.

Table 3

Adsorption energies ($E_{ads}$) of H, C, N, O, and CO, as estimated by the CCSD(T)/cc-pVTZ level of theory. The minimum and maximum adsorption energies in our calculations are shown.

| Species | $E_{ads}$ (K) | |
| --- | --- | --- |
| | site A | site B |
| H | 60 | 60 |
| C | 350–520 | > 20,000 |
| N | 240–260 | 230–240 |
| O | 290–420 | 470–500 |
| CO | 290–500 | N/A[a] |

Note.

[a] Geometric optimization did not converge.

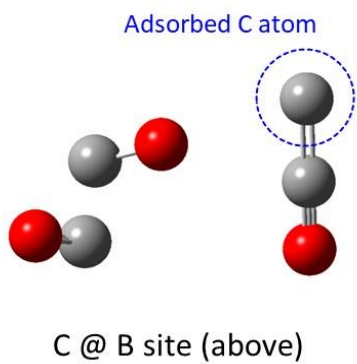

C @ B site (above)

Figure 18. Adsorbed structure of the C atom at the B site (from above) of the α-CO (111) surface. The position of the adsorbed C atom was only optimized.

*5.2. Non-thermal desorption of molecules: Photodesorption and chemical desorption*

Here, we briefly discuss the effect of the morphology of icy grains proposed in this work upon the non-thermal desorption (particularly the photodesorption and chemical desorption) of molecules. It is now widely accepted that photodesorption plays an important role in regulating the partitioning of molecules between gas and ice in molecular clouds/cores (e.g., Caselli et al. 2012) and in the cold outer regions of proto-planetary disks (e.g., Hogerheijde et al. 2011; Öberg et al. 2015), where the dust temperature is low (<20 K). Laboratory experiments and molecular-dynamics simulations have shown that photodesorption can desorb molecules only from the surface of the ice (i.e., the topmost several monolayers) (e.g., Öberg et al. 2009a; Arasa et al. 2015); therefore, the morphology of icy grains significantly affects the efficiency of molecular photodesorption. If icy grains have a classical onion-like (or layered) structure, $H_2O$ photodesorption would be hindered once CO adsorbs on top of the $H_2O$ ice layers, burying them (see Sect. 4.3). On the other hand, if icy grains have the morphology proposed in this work, $H_2O$ photodesorption would be efficient even after significant CO adsorption occurs on top of the $H_2O$ ice layers. The latter case may be more consistent with the detection of $H_2O$ vapor in the center of the pre-stellar core L1544, where significant CO freeze-out has already occurred (Caselli et al. 2012). Indeed, Caselli et al. (2012) successfully reproduced the $H_2O$ emission line observed in L1544 using their radiative-transfer and chemical models with $H_2O$ photodesorption, the rate of which was calculated assuming that the surfaces of the ice mantles are fully covered by $H_2O$ (see also Keto et al. 2014); this assumption neglects the fact that interstellar ices contain significant fractions of other molecules, such as CO; our proposed morphology of icy grains suggests that this assumption is reasonable. $H_2O$ photodesorption experiments, whereby CO is deposited on top of $H_2O$ ice, should be conducted.

The importance of chemical desorption (or reactive desorption) in the partitioning of molecules between the gas and solid phases in molecular clouds has also been recognized in chemical modeling studies (Garrod et al. 2007; Cazaux et al. 2010; Wakelam et al. 2017); recently, it has been experimentally demonstrated that various kinds of molecules can desorb upon their formation on icy surfaces at low temperatures (Minissale et al. 2016; Oba et al. 2018; Chuang et al. 2018; Nguyen et al. 2020). Additionally, the desorption efficiency can vary depending on the icy surfaces' structures (Oba et al. 2018, 2019). Although there is no data concerning the chemical-desorption efficiency of molecules from CO-covered surfaces, we expect that it will differ from the desorption efficiency from $H_2O$ surfaces, whose differences need to be precisely incorporated into future modeling studies.

### 5.3. Collision and sticking

In general, the adhesion force between grains in contact is primarily determined by the composition and morphology of their contact surface; therefore, grains with highly heterogeneous surface compositions and morphologies (as depicted in Fig. 16) may stick significantly differently from ellipsoidal and uniformly layered grains. For example, simple grain models assuming layered mantles of various volatiles (e.g., Musiolik et al. 2016; Pinilla et al. 2017; Okuzumi & Tazaki 2019) predict that the stickiness of grains in a disk is uniquely determined by their location relative to the relevant snow lines; however, our experimental results offer a more complex picture in which grains can have heterogeneous surface compositions and the combination of materials constituting their contact surface is not unique (consider combinations between icy grains depicted in the bottom of Fig. 16). The sticking efficiency of two grains in a disk cannot be uniquely predicted as a function of their location and can instead vary depending on their relative orientations upon collision.

Similarly, the irregular surface morphologies originated from the existence of single crystals on grains inevitably introduce some randomness to their sticking efficiencies. Models of proto-planetary dust growth conventionally assume that there exists a critical collision velocity, $v_c$, below which two grains can stick. This is a reasonable assumption for uniformly layered grains; however, no well-defined threshold velocity is likely to exist for crystalline particles with facets and edges. While a contact between faces would apply a large contact area and hence a large binding energy, the opposite would apply for a contact between edges. In fact, Poppe et al. (2000) have already reported a lack of a well-defined maximum sticking velocity for irregularly shaped grains.

However, the above discussion does not necessarily mean that *aggregates* of uniformly layered or irregular, heterogenous grains should have significantly different sticking properties, because an aggregate generally consists of a number of contact surfaces, and their bulk mechanical properties may be determined by the average properties of the surfaces. Further experiments and modeling efforts are needed to better understand the statistical sticking properties of heterogeneous grains and their aggregates.

### 5.4. Sintering

The collisional outcome of sintered aggregates differs substantially from that of non-sintered aggregates (Sirono & Ueno 2017), because the contact between constituent grains is strengthened by sintering. When one considers possible combinations of contacts between icy grains depicted in the bottom of Fig. 16, sintering should proceed if both sides of a contact are made of the same kind of ice, as discussed in Sirono (2011); here, the strength of a contact increases by sintering. If the composing materials of two sides of a contact differ, sintering depends upon the surface energies of the two composing materials. For example, sintering of $CO_2$ ice on $H_2O$ ice is more likely to occur than sintering of CO ice on $H_2O$, due to the difference in the wettabilities of the pair of ices. At this stage, it is difficult to predict whether sintering proceeds or not for a particular pair of ices. The strengths of contacts inside an aggregate would vary; although it is clear that the collisional-growth efficiency of uniformly sintered aggregate is lowered (Sirono & Ueno 2017), the collisional outcome of non-uniformly sintered aggregate is not clear and need to be studied.

Acknowledgments

This work was supported by the Ministry of Education, Culture, Sports, Science, and Technology Grants-in-Aid for Scientific Research (KAKENHI Grant Numbers JP25108002, JP18H05438, JP18H05441, JP20H04676, JP20H05849) and Japan Society for the Promotion of Science Grants-in-Aid for Scientific Research (JSPS KAKENHI Grant Numbers JP17H01103, JP17H06087, JP18H01262, JP19K03926, JP19K03941, JP20H00182, JP20H00205, JP21H01139, JP21H04501).

L174